\documentstyle[psfig,epsfig,subfigure,rotating,lscape]{l-aa}

\def \rsun {\ifmmode$R$_{\odot}\else R$_{\odot}$\fi}

\def \hcm {\hbox {\ifmmode $ atoms cm$^{-2}\else atoms cm$^{-2}$\fi}}
\def\approxgt{\mathrel{\hbox{\rlap{\lower.55ex \hbox {$\sim$}}
        \kern-.3em \raise.4ex \hbox{$>$}}}}
\def\approxlt{\mathrel{\hbox{\rlap{\lower.55ex \hbox {$\sim$}}
\kern-.3em \raise.4ex \hbox{$<$}}}}
\include{epsf}

\begin{document}
 

\thesaurus{04.01.1, 04.03.1, 13.25.3}

\title{The EXOSAT medium-energy slew survey catalog}

\author{A.P. Reynolds  \and A.N. Parmar\  \and P.J. 
Hakala\  
\and A.M.T. Pollock\  \and O.R. Williams\ 
\and A. Peacock\  \and B.G.~Taylor\ }

\institute{Astrophysics Division, Space Science Department of ESA,
ESTEC, P.O. Box 299, 2200 AG Noordwijk, The Netherlands}

\offprints{A.P. Reynolds: areynold@astro.estec .esa.nl}
\date{Received 17 March 1998; accepted 24 July 1998}

\maketitle

\markboth{A.P. Reynolds et al.}{EXOSAT ME slew survey}

\begin{abstract}
 
We present a catalog of X-ray sources observed during slew maneuvers
by the Medium Energy Detector Array onboard the
EXOSAT Observatory. 
The EXOSAT Medium Energy slew-survey catalog (EXMS) provides
a unique record of the 1--8 keV X-ray sky
between 1983 and 1986.
98\% of the sky was observed, with 85\,\%
receiving an exposure of $>$60~s. 1210 sources were detected.
By comparing these source positions with other
catalogs, identifications are given for 992 detections 
(82\% of the sample).
These identifications consist of 250 distinct
objects, including 95 different X-ray binary systems, 
and 14 different AGN. 
A further 58 detections
have multiple candidates, while 160
detections remain unidentified.
Collimator transmission corrected 1--8~keV 
count rates are given for the identified sources, together with
raw count rates for the other detections.
The construction of the EXMS and the checks
performed to ensure the validity of the derived source properties are 
discussed. A publically available version of this catalog is 
maintained on the EXOSAT database and archive system 
(telnet://xray@exosat.estec.esa.nl).

\keywords{astronomical data bases: miscellaneous -- catalogs -- X-rays: 
general}
\end{abstract}

\section{Introduction}

Sky surveys are of particular importance in high energy astronomy,
where many sources exhibit irregular long-term variability which
cannot be conveniently monitored by pointed observations.
Observations conducted while 
maneuvering between targets can provide a substantial bonus
to the scientific return of pointed missions (e.g., Elvis et al. 1992).
Such observations are complementary to the dedicated all-sky surveys 
conducted by scanning instruments such as the {\it Uhuru},
Ariel-V, HEAO A-1, and ROSAT bright source catalogs (Forman et al. 1978; 
Warwick et al. 1981; McHardy et al. 1981; Wood et al. 1984; Voges et al. 1996).
Here we report the second major X-ray slew survey, 
derived from observations made
by the European Space Agency's EXOSAT X-ray astronomy satellite 
(White \& Peacock 1988).

EXOSAT  performed
1780 pointed observations of a wide variety of objects
between 1983 June and 1986 April. 
The 90~hr orbit had an apogee of 190,000~km and 
perigee of 350~km, with the science payload operated when the
satellite was above the Earth's radiation belts at 50,000~km.
 This allowed uninterrupted 
observations of up to 76~hr duration. 
The satellite was three axis stabilized, and at any given time about
half the sky could be viewed.
X-ray sources were simultaneously observed with up to 4 coaligned 
instruments.
Two Channel Multiplier Array detectors (CMA; de Korte et al. 1981)
each at the focus of an
X-ray mirror provided images in the low-energy (0.04--2.0~keV) energy range,
while the Medium Energy Detector Array (ME; Turner et al. 1981)
and the Gas Scintillation Proportional Counter (GSPC; Peacock et al. 1981)
covered the 1--50~keV and 2--35~keV energy ranges, respectively. 
In addition to the the pointed observations, a series
of slews along parts of the galactic plane were performed
(Warwick et al. 1985, 1988).

When EXOSAT manoeuvred between targets the ME and GSPC instruments were 
usually operated in order to search for new sources, monitor known ones, 
and to measure background counting rates.
The relatively slow manoeuvre rates of EXOSAT (either 42, 85, or 
170$^\circ$~hr$^{-1}$), together with a good knowledge of the pointing
direction during slews,
allows the construction of a catalog with high sky-coverage and 
sensitivity.
Slew manoeuvres were usually performed in three
stages or legs, rather than along the 
great circles directly between sources.
First there would be a 
slew to place the instruments' pointing axis 90$^{\circ}$ from the Sun
(at a $\beta$ angle of $90^{\circ}$), followed by a slew along the 
$\beta = 90^{\circ}$ line and then a final 
slew off the $\beta = 90^\circ$ line to the new pointing position. 
Apart from maximizing the efficiency of solar power collection,
this procedure resulted in greater sky coverage
than if the slews were along connecting great circles. 
During EXOSAT operations the ME slew data were routinely checked for the
presence of X-ray sources. This led to the discovery of four 
previously unknown X-ray sources, all of which were found to be
X-ray binaries (see White \& Peacock 1988).

\section{The Medium Energy Detector Array}
\label{sect:me}

\subsection{The detectors}
\label{subsect:detectors}

\begin{figure}
\begin{center}
\epsfig{file=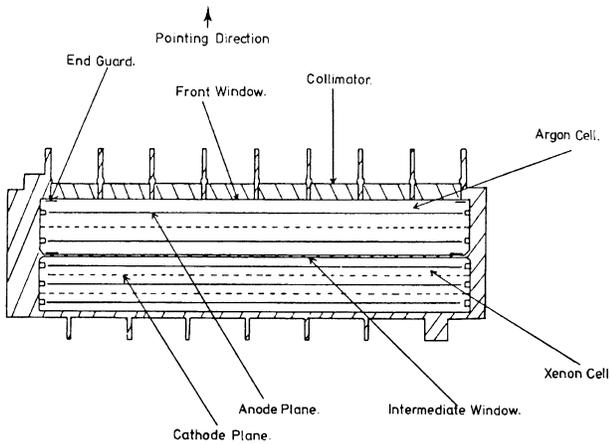, 
width=3.5in, angle=0}
\caption{Cross-section of an ME detector}
\label{fig:me_detectors}
\end{center}
\end{figure}

The performance of the ME is summarized in Table~\ref{tab:me}.
The ME comprised 8 individual detectors, each of which
consisted of an Ar/CO$_{2}$ and a Xe/CO$_{2}$ 
gas filled multi-wire proportional counter separated by a 1.5~mm thick
Be intermediate window (Fig.~\ref{fig:me_detectors}). 
An X-ray collimator, made from lead-glass microchannel plates
was mounted in front of each detector.
Anticoincidence and pulse rise time techniques were used to
reduce the particle background.
The ME operated well throughout the mission with  
one of the detectors failing on 1985 August 20. Problems with
detector breakdown at the start of the mission were solved by
operating the Ar counters at a lower overall gain setting. This resulted
in pulse-height analyzer (PHA) channel 128 
corresponding to an energy of $\sim$50~keV.
 
\subsection{Background}
\label{subsect:background}

The ME background was usually stable with time and dominated by
particle-induced events from the solar wind and 
events from the radioactive lines of residual
Plutonium in the Be windows and detector bodies. The contribution
of the extragalactic X-ray background was $<$1\% of the total Ar
background counting rate.
After anticoincidence 
rejection of particle-induced events the 
typical 1--8~keV Ar background count
rates was 3.8~$^{-1}$~detector$^{-1}$.
Occasional background flares occurred simultaneously
in some or all of the detectors and were caused 
by enhancements in the solar wind.
For normal observations longer than $\sim$5000~s,
the ME was limited by systematic effects in the
background subtraction to detections of 
$\approxgt$$0.5\times 10^{-11}$~erg~cm$^{-2}$~s$^{-1}$.

\subsection{Field of view}
\label{subsect:fov}

\begin{figure}
\begin{center}
\epsfig{file=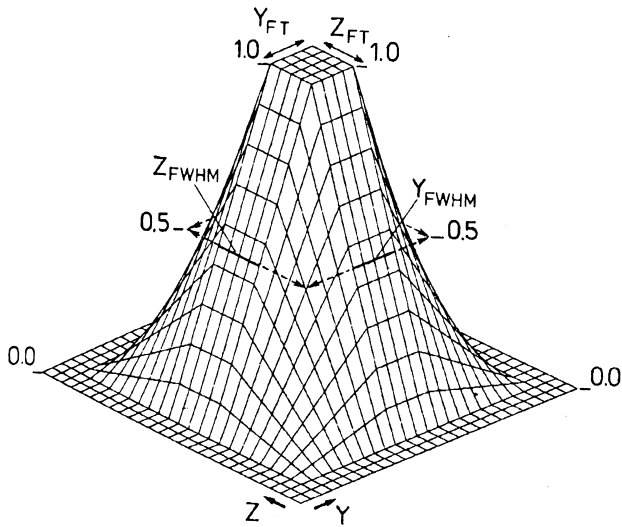, 
width=3.5in, angle=0}
\caption{A schematic model of an idealized ME collimator response 
(Kuulkers 1995). Slew directions were always parallel to the Y or
Z axes. The flat top (Y$_FT$, Z$_FT$) to the collimator response 
had approximately equal sides
of length 7 arc mins. The numbers indicate the relative transmission}
\label{fig:collimator}
\end{center}
\end{figure}

The ME field of view (FOV) was defined by collimators which had
a rectangular aperture with an average full-width half-maximum (FWHM)
of 45$'$ and a flat top of $\sim$7$'$ (Fig.~\ref{fig:collimator}).
Details are to be found in Gottwald (1984) and Kuulkers (1995). 
Figure~\ref{fig:collimator} shows a labeled schematic model of the 
collimator response.
The 8 detectors were grouped into four pairs or ``quadrants'', each
of which could be offset by up to 120$'$
from the aligned position.
For most targets, two quadrants (one half of the ME) would be aligned and 
pointed at the
target, while the remaining quadrants would be offset and
pointed at two adjacent regions of sky in order to monitor the background
counting rate. 
For bright sources, where background subtraction is not so critical, 
the ME could be operated with all four quadrants observing the source.
The orientations of the quadrants are stored in the spacecraft
pointing files, which were updated every 60~s.

\subsection{Additional slew observations}
\label{subsect:warwick}

As well as the slew manoevers required to move between
scheduled pointings, EXOSAT performed a series of scans
along parts of the galactic plane as part of the scientific program.
The results are presented in Warwick et al. (1985, 1988) and are not
included in the EXMS.

\begin{table}
\caption[]{Properties of the EXOSAT ME}
\begin{flushleft}
\begin{tabular}{ll}
\hline
Geometric Area (8 detectors)      & 1600~cm$^{2}$\\
Field of view (FWHM)              & $\sim$$45' \times 45'$\\
Ar counter energy range           & 1--20 keV\\
Xe counter energy range           & 5--50 keV\\
Number of PHA channels            & 128 + 128 (Ar and Xe) \\ 
Energy Resolution (Ar)            & 49/E(keV)$^{0.5}$ \,\% FWHM\\
1--8 keV Background (8 detectors) & 21 counts~s$^{-1}$ \\ 
\hline
\end{tabular}
\end{flushleft}
\label{tab:me}
\end{table}

\section{The EXOSAT ME slew survey catalog (EXMS)}
\label{sect:mess}

\subsection{Data selection}
\label{subsect:data}

As part of the routine data processing during
the EXOSAT post-operational phase, slew lightcurves 
covering the energy ranges 1--8~keV and 10--18~keV were systematically 
produced. 
If data were available for individual detectors or quadrants, then three
lightcurves were produced for each energy range and time interval;
one for the two aligned quadrants and one for each of the offset quadrants 
(since they
pointed in different directions on the sky). If data were only
available for each half-array, then separate
lightcurves for the aligned and offset quadrants were produced for 
each energy range.
In this case, a source detected in an offset
quadrant could not always be unambiguously located on the sky, since
the counts may have arisen from one of two separate locations.
In the cases when a single
data stream covering the entire detector was selected, only a
single set of lightcurves covering both the aligned and offset quadrants was
produced. The time taken for a source to pass through a FOV 
is dependent on the slew rate, which was nominally
42$^{\circ}$\,hr$^{-1}$, with two faster
speeds of 85 and 170 $^{\circ}$\,hr$^{-1}$ also used. At the slowest
rate, a source would take 135\,s to pass completely through the FOV,
from zero response to zero response.
The 1--8~keV energy range was chosen since this is where the residual
background in the ME was lowest and the sensitivity highest for a source
with a typical X-ray spectrum. The 10--18~keV energy range was chosen to
sample the solar particle induced background
counting rate; the contribution of most X-ray sources in this energy range
being small. The slew lightcurves were accumulated with a binning time
of 10~s. This was chosen to allow a uniform set of lightcurves to be
produced,
since all the ME data configurations provided spectra with 10~s or faster time
resolution, while allowing the collimator profile to be adequately sampled
while slewing.  

\subsection{Exposure map}
\label{subsect:exposure}

\begin{figure*}
\begin{center}
\epsfig{file=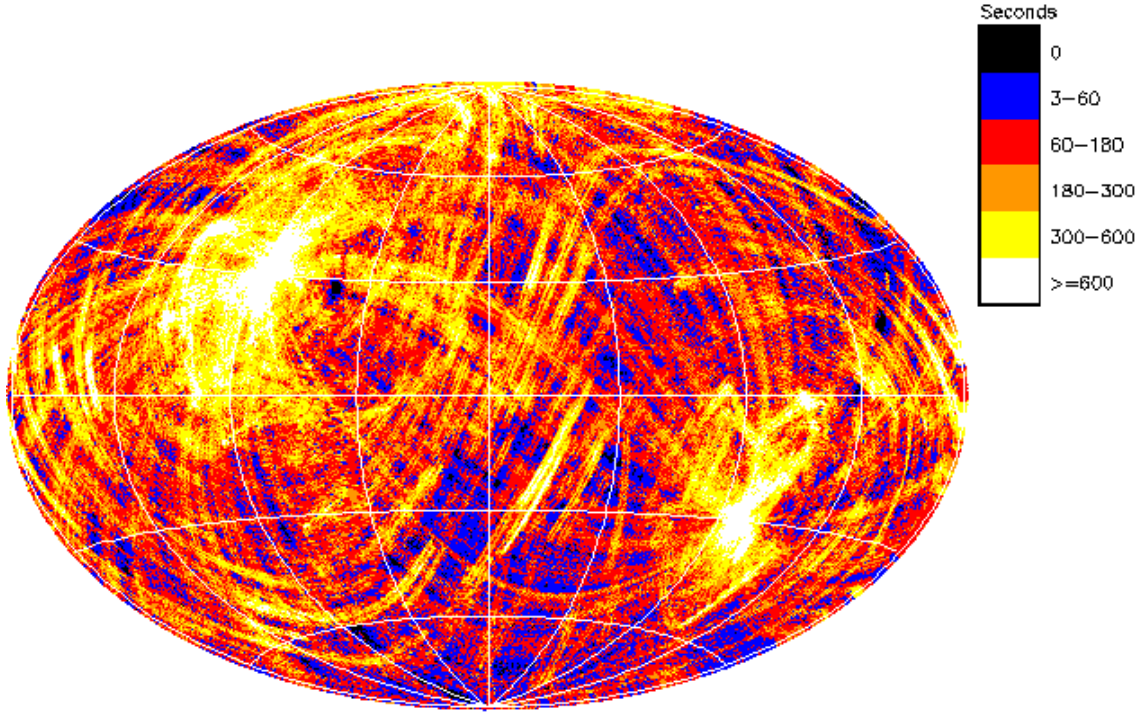, width=6.0in, angle=0.0}
\caption{Exposure map for the complete EXMS catalog in galactic
coordinates. 
Areas which were
never slewed over, or had exposures of $<$3~s are shaded black. All 
colored regions received an exposure of $\ge$3~s. Areas colored 
white received exposures
$>$600~s. 98\% of the sky was slewed over at least
once and 85\% of the sky received an exposure of $>$60~s}
\label{fig:sky_coverage}
\end{center}
\end{figure*}

Data from a total of 1765 slews, $>$99\,\% of the total were processed. 
Figure~\ref{fig:sky_coverage} 
is an exposure map 
of the data used to construct this catalog. The map
was generated by first calculating the region of sky swept out by each quadrant
on the celestial sphere, using the positional information in the
spacecraft pointing files.
The sky was then divided into cells of width
0.25$^{\circ}$, 
and each cell was assigned an effective exposure, 
dependent on the slew rate and the distance of the cell from
the centroid of the collimator. Individual cells were often slewed over
repeatedly, especially near the ecliptic poles, and the effective
exposures were updated for each subsequent slew over a given cell.
 These values were then converted
into contours and plotted in a Hammer-Aitoff projection
using Galactic coordinates. The efficiency of sky coverage was
high: 98\% of the sky was slewed over at least 
once, while 85\% received an exposure of $>$60~s.
The few regions of the sky which were not slewed
over at all, or had exposures $<$3~s, or where there is no
pointing data,
are indicated in black. All other areas received exposures between
3~s and 6000~s. This compares favorably with the exposure
obtained by the {\it Einstein} slew survey, where useful sensitivity
was only obtained for 50\% of the sky (Elvis et al. 1992). As in the 
{\it Einstein} slew survey,
the
areas of sky which received the highest exposure are near the ecliptic poles,
but unlike the latter
the Galactic plane is also well-sampled, since there was
no requirement to avoid slewing across bright sources. This means that
the slew survey (hereafter EXMS)
is expected to be rich in compact X-ray sources.

\begin{figure*}
\begin{center}
\epsfig{file=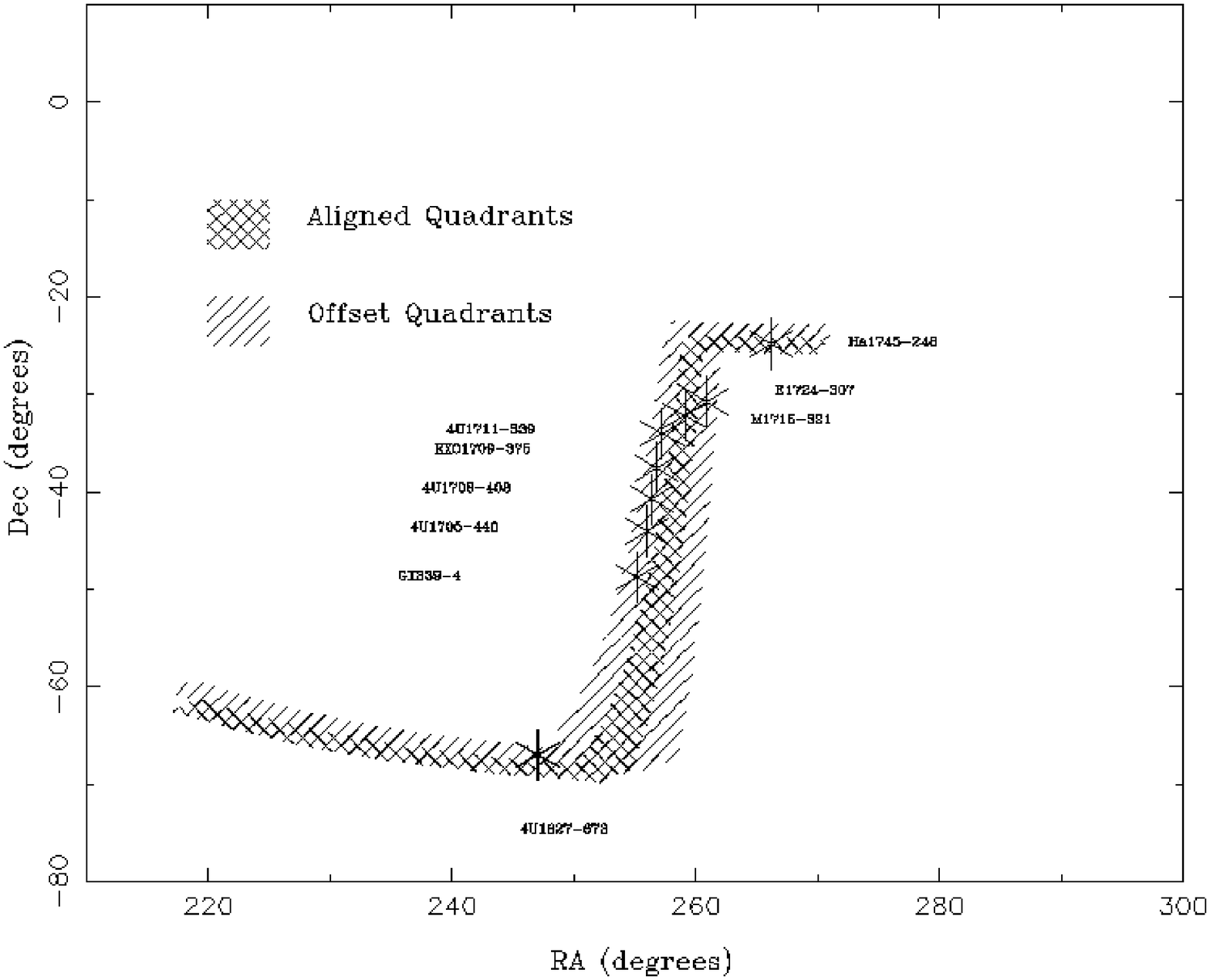, width=5.0in, angle=0}
\caption{The ``footprint'' of a representative three-legged slew showing
the area of sky observed. The hatched and cross-hatched areas show the
area covered by the offset and aligned quadrants, respectively. The 
start of the slew was in the lower left corner. The 
variation in the width of the footprint
results from 
the projection onto equatorial coordinates.
The positions of
the sources identified in Fig.~\ref{fig:lightcurve1} are marked with
stars} 
\label{fig:footprint}
\end{center}
\end{figure*}

A schematic diagram of a typical three-legged slew is shown in
Fig.~\ref{fig:slew_schematic}. The direction of slew is always
parallel to one edge of the rectangular FOV. In Fig.~\ref{fig:footprint}
the region of the sky swept out by a real three-legged slew is shown. 
In the first, second, and third legs, the FOVs swept 
out two, three and two strips of sky, respectively. 
 The differing paths swept out by the 
aligned and offset quadrants are shown, together with the positions
of the eight detected sources.

\begin{figure}
\begin{center}
\epsfig{file=slew_schematic.ps, width=2.2in, angle=-90}
\caption{Schematic diagram of a three-legged slew, with the
four quadrants of the ME arranged in the usual combination
of two aligned and two offset. The squares indicate the FOVs
of each quadrant and are not to scale. Slew directions were
always parallel to one edge of the FOV} 
\label{fig:slew_schematic}
\end{center}
\end{figure}

\subsection{Source detection}
\label{subsect:detection}

To illustrate the quality and type of data available in the EXMS, 
Fig.~\ref{fig:lightcurve1} shows representative 1--8~keV lightcurves for
a slew across a crowded region of sky 
(the same slew as illustrated in Fig.~\ref{fig:footprint}). 
The three panels show
lightcurves for the aligned and two offset quadrants.
One source (4U\thinspace1627$-$673) is seen in the two offset quadrants at
different times (separated by 240~s). All count rates are
normalized to s$^{-1}$~half$^{-1}$ (4 detectors), but no
correction for collimator losses has been applied. The
lightcurves are not background subtracted and the nominal background
counting rate of $\sim$15~s$^{-1}$~half$^{-1}$ is visible whenever
there are no sources in the FOV.
The 3 slew legs can be seen
in the lightcurves
since there are missing data between the slew legs (just after 16~hr and
just before 16.6~hr). This is due to data from contiguous
intervals of less than 5 minutes duration not being stored on EXOSAT 
data tapes.

The slew lightcurves were systematically searched for sources passing
through the FOV by modeling the observed time-dependent 
1--8 keV count rate, ${\rm R_{1-8}}$, by:

${\rm  R_{1-8}(T) = S H(T-T_{s}) + k R_{10-18}(T)} $

The source term depends on the source count rate, S, the time of
transit through the collimator center, ${\rm T_{s}}$, and the 
collimator profile
${\rm H(T-T_{s}})$ determined by the slew rate. The 1--8~keV background 
count rate was
assumed to be linearly related through a constant, k,
to the 10--18~keV count rate, ${\rm R_{10-18}}$.
This was verified by inspection of the lightcurves which showed that only
sources with peak count rates $\approxgt$100~s$^{-1}$ were detected in the 10--18~keV
energy range.
This approach is successful in excluding the type of background events
illustrated in the upper panel of Fig.~\ref{fig:lightcurve1}, 
in which short-term increases in
the solar particle flux can mimic the intensity profile of an 
Z-ray source.

\begin{figure*}
\begin{center}
\epsfig{file=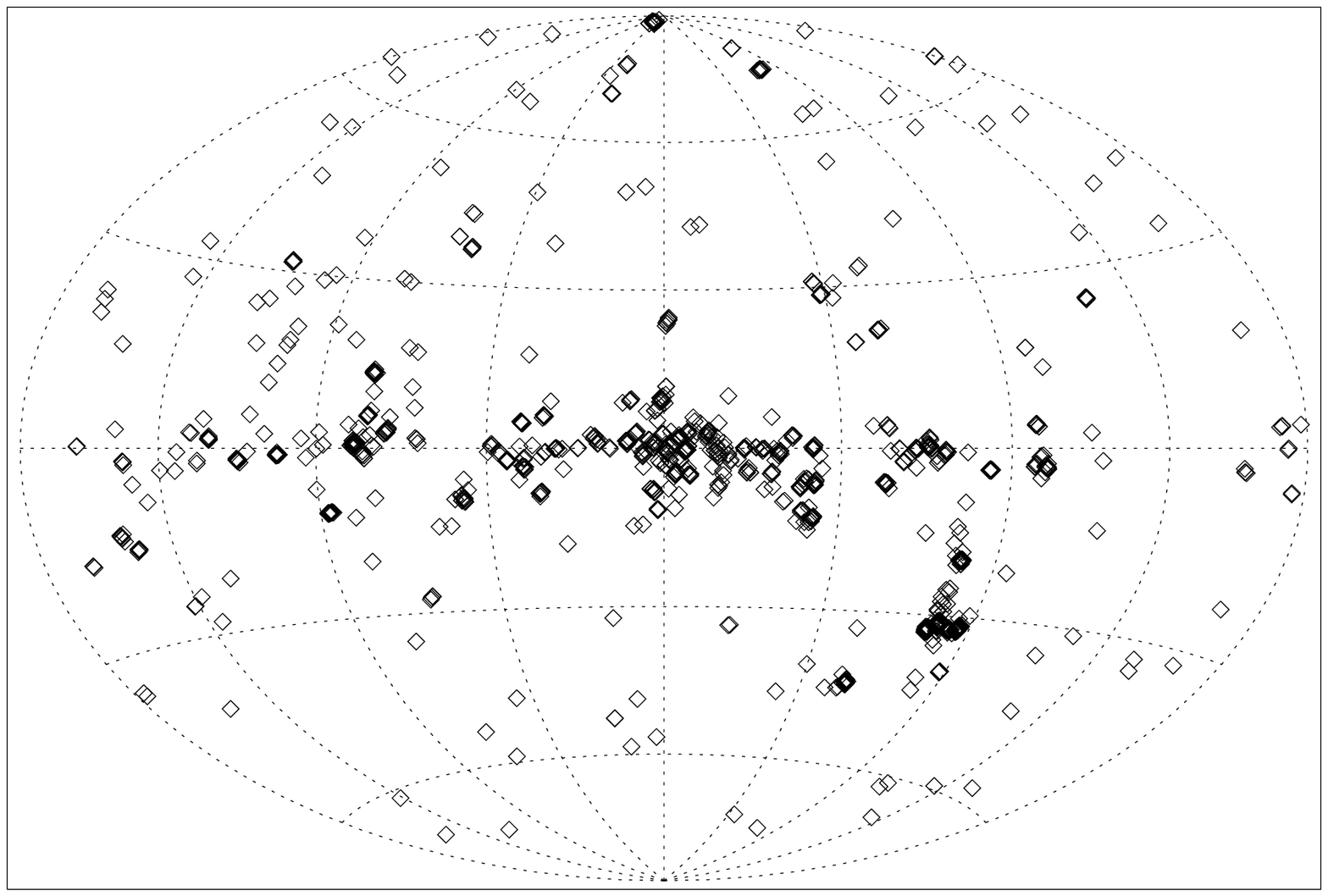,bbllx=0pt,bblly=360pt,bburx=1028pt,bbury=720pt,
clip,width=7.0in, angle=0}
\caption{The 1210 EXMS detections in galactic coordinates. The
predominance of galactic plane and Magellanic cloud sources is 
evident}
\label{fig:detection_map}
\end{center}
\end{figure*}

For a fixed ${\rm T_{s}}$ best-fit parameters S and k were 
determined by maximum
likelihood. Candidate detections were initially identified in a 
coarse search in
${\rm T_{s}}$, after which parameters and their uncertainties were precisely 
calculated. The transit of some strong
sources can be located to $<$0.1~s. 
Occasionally, it was necessary
to fit two sources simultaneously. 
The limiting 1--8~keV sensitivity, when no confusing sources are present,
is estimated to be $3\times 10^{-11}$~erg~cm$^{-2}$~s$^{-1}$ at the
slowest slew rate.
The source best-fit
locations and error boxes were reconstructed using the values
of ${\rm T_{s}}$ and ${\rm \Delta{T_{s}}}$
by parabolic interpolation from the positions contained in the pointing files.
A total of 1210 detections were made in this way and are
shown in galactic 
coordinates in Fig.~\ref{fig:detection_map}.

If the fit to the collimator profile
is formally acceptable, the time of maximum intensity 
corresponds to the position of the quadrant when the source
passed through its midline. 
Hence, source positions can be
well localized in the direction of slew, but the
position perpendicular to the slew axis is constrained only by the
width of the FOV, unless assumptions about the source intensity
are made.
This means that, for the
majority of sources,
the uncertainty regions are narrow in the slew
direction and much broader perpendicular to it.

\begin{figure}
\begin{center}
\epsfig{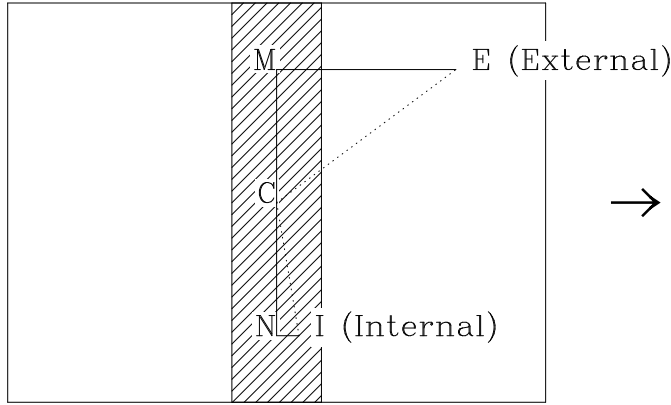}
\caption{Schematic diagram of a formal EXMS uncertainty
region, shown as a hatched rectangle within the rectangular FOV.
The arrow shows the slew direction.
The source positions I and E are internal and external
to the uncertainty region, respectively. 
C is the uncertainty region centroid. 
EC and IC are the distances between
sources E and I and the centroid. 
EM and IN are the parallel, or vector, 
offsets in the direction parallel to the
slew.
CM and CN are the distances which determine the correction
for collimator transmission} 
\label{fig:errorbox}
\end{center}
\end{figure}

\begin{figure}
\begin{center}
\epsfig{file=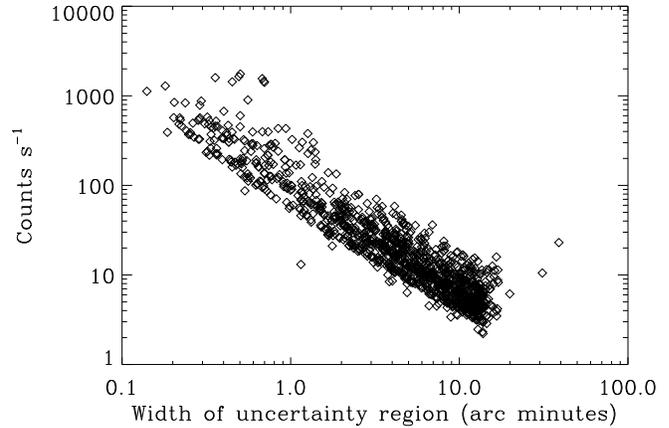, width=3.5in, angle=0}
\caption{Plot of the dimensions of the short sides of 1210
uncertainty regions against source count rate.
The clustering of the data points into
three distinct lines is a consequence of the
three different slew rates, the slowest of which allowed
sources to be localized most precisely}
\label{fig:dimensions}
\end{center}
\end{figure}

The mean length of the narrow side of the 1210 uncertainty regions is
(6.30 $\pm$ 0.12)$'$; (all errors are given at 68\% confidence).
In Fig.~\ref{fig:errorbox} a schematic diagram of a typical EXMS
uncertainty region is shown defining the different offsets which are
referred to later. 
In Fig.~\ref{fig:dimensions}
the lengths of the narrow sides are plotted against uncorrected source
counts for the 1210 detections. This demonstrates 
that the brightest detections tend to have the
narrowest uncertainty regions, as expected. 
Since these regions are constructed
under the assumption of a constant source intensity, it is to be
expected that some identifications will lie outside their
formal uncertainty region (see Fig.~\ref{fig:errorbox}).
Also seen in Fig.~\ref{fig:dimensions} is the effect of
the different slew rates on the width of the uncertainty region
for a given count rate, leading to three distinct clusterings
of points, superimposed on the scatter caused mainly by 
source variability.

\begin{figure*}
\begin{center}
\psfig{file=quad12_lc.ps, width=4.0in, height=2.8in, angle=0}
\psfig{file=quad3_lc.ps, width=4.0in, height=2.8in, angle=0}
\psfig{file=quad4_lc.ps, width=4.0in, height=2.8in, angle=0}
\caption{Lightcurves obtained during a slew across a crowded
region of sky, 
obtained from the two aligned quadrants (upper panel)
and the two offset quadrants (middle and lower panels). No correction
for collimator losses has been applied.
Source identifications are given.
The count enhancement at around 16.4~hr in the aligned quadrants
is due to increased solar activity.
The slew rates were 
42$^{\circ}$\,hr$^{-1}$, 85$^{\circ}$\,hr$^{-1}$ and 42$^{\circ}$\,hr$^{-1}$ 
during the first, second, and third slew legs, respectively. The intervals
of missing data correspond to the breaks between slew legs.
Note that 4U\,1627$-$673 was detected in both offset quadrants
separated by 240~s}
\label{fig:lightcurve1}
\end{center}
\end{figure*}

\subsection{Source indentification}  
\label{subsect:ident}

Where possible, sources responsible for the 1210 detections have
been identified by correlation against previous
catalogs. The procedure is to determine all the sources
which fall within the 99.7\%
confidence uncertainty region
for a particular detection. This may include entries from
more than one catalog.
A preferred identification is
then made using a hierarchical selection scheme. The following
catalogs were used to generate possible identifications,
and are listed in hierarchical order:

\begin{itemize}

\item XRBCAT: A catalog of X-ray Binaries derived from 
van Paradijs (1995). 

\item RITTER: A catalog of cataclysmic variables, X-ray binaries and
related objects, derived from Ritter (1990).

\item VSTARS: The General Catalog of Variable Stars, derived from the
69th name list of variable stars (Kholopov et al. 1989).

\item XRAY: A master catalog containing selected parameters from all
X-ray catalogs present on the 
online system, including HEAO-1, {\it Einstein}, EXOSAT and ROSAT source
catalogs.

\item VERON96: Derived from the 7th edition of
the Catalog of Quasars and Active Galactic Nuclei by 
Veron-Cetty \& Veron (1996).

\item RADIO: A master catalog containing selected
parameters from several radio source catalogs.

\end{itemize}

At the end of this search, a given detection would either
have produced no identification, a single identification, or
multiple candidates. In
order to ensure that these identifications were always
known X-ray emitters, a further search was made against the
ROSAT All-Sky Survey Bright Source catalog (RASSBSC; Voges et al. 1996; 1998),
which had not been incorporated into the XRAY
catalog at the time of the initial search. The RASSBSC was
used to validate all proposed identifications and to search
for any additional candidates
which were not present in the other catalogs.
Consequently, after cross-checking against the RASSBSC, many
candidate identifications derived from the VSTAR, RADIO and
VERON96 catalogs were discarded. 
Known transient X-ray
sources, and extended X-ray objects (which do not appear in the
RASSBSC) are retained. 

The sources found in the above searches are all within
the formal uncertainty regions. However, apart
from these ``internal'' sources, attention was also paid to any 
bright, time variable and/or
extended sources whose derived positions were outside the formal 
uncertainty regions, but which still passed within the FOV during a slew.
Such sources, typically X-ray binaries or
supernova remnants, may well have incorrectly determined
uncertainty regions due to time-variability, or their
extended nature.
In practise, whenever this occured, 
there was little doubt that the
detection was due to the ``external'' source, since there was
either no candidate in the uncertainty region, or a source
not known to be a strong X-ray emitter.

Some sources are considered confused; i.e., while
it is likely that the majority of counts originate from the
object proposed as the identification, the RASSBSC 
indicates that at least one fainter source 
was present in the uncertainty region. The fainter source
{\it may} therefore
contribute to the signal.
The criterion used for classifying a source as confused
is that the brightest RASSBSC source
must be at least four times brighter than any other source 
within the uncertainty
region. Where this is not the case (ie where the RASSBSC
sources differ by less than a factor of four in brightness)
no single identification is proposed. In this case we 
assume that the brightest RASSBSC
source (based on 0.1--2.4~keV ROSAT Position Sensitive Proportional Counter
countrates)
is also the brightest in the 1--8 keV EXMS range.


\begin{table}
\caption[]{Counterpart statistics and flag values}
\label{tab:quality_flags}
\begin{flushleft}
\begin{tabular}{lllrll}
\noalign {\smallskip}
\hline
\noalign {\smallskip}
Flag & N${\rm _{samp}}$ & \% & N${\rm _{int}}$$^{a}$ & 
N${\rm _{ext}}$$^{b}$ \
& Comment \\
\noalign {\smallskip}
\hline
\noalign {\smallskip}
4 & 617 & 51\,\%  & 1 & 0 & Internal \\
3 & 355 & 29\,\%   & 0 & 1 & External \\
2 & 20  & 2\,\% & \llap{$>$}1 & 0 & Confused \\ 
1 & 58  & 5\,\% & \llap{$>$1} & 0 & Multiple\\
0 & 160 & 13\,\%   & 0 & 0 & Unidentified \\
\noalign{\smallskip}
\hline
\end{tabular}
\end{flushleft}
$^{a}$Number of sources within the uncertainty region.

$^{b}$Number of sources outside the uncertainty region,\\ 
but within the FOV.

\end{table}

Flags have been assigned to all entries to distinguish the type
of identification; they may also be considered as approximate quality
flags.
This
scheme is summarized in Table~\ref{tab:quality_flags}, together
with the numbers and percentages of sources in
each category. 51\% of the entries in the slew
survey are identified with confidence - ``internal'' sources where
the most likely X-ray source also lies within the uncertainty
region. A further 29\% of entries consist of identified
sources which lie outside their formal uncertainty regions. Although these
identifications are considered secure, the derived count rates
are likely to contain larger errors than the internal sample,
due to the error in the initial fit to the lightcurve (this
problem is particularly severe for X-ray pulsars where the
duty cycle of the pulsations may be $\approxlt$50\%). The
mean reduced $\chi$$^{2}$ for the internal sample is 1.3, 
whereas that for the external sources it is 48.
The majority of these identifications are X-ray binaries. Hence,
82\% of the entries
are given single identifications, similar to the identified
fraction in the {\it Einstein}\, slew survey. A further 2\%
of sources are ``confused'' - there is a fainter
source (at least {\it four times fainter}, based on RASSBSC count rates) 
present in the uncertainty bound.

A further 5\% of
entries consist of cases where multiple sources of similar
brightness lie within the formal uncertainty region.
13\,\% of the entries cannot be identified with any known X-ray emitters,
either because there are no X-ray emitters in the vicinity or there
is significant doubt about the likely identification.
In the case of the 82\% of entries where one identification is
proposed (sources with Flags
of 2, 3, and 4), the count rates given in Table~3 are corrected
for collimator transmission. 
This is achieved by calculating the
offset of the source (based on its catalog coordinates) from
the centroid of the detector (see line CM in 
Fig.~\ref{fig:errorbox}).
The correction is proportional
to this offset, except for the small ``flat-top'' at the center of the FOV
(see Fig.~\ref{fig:collimator}).
In addition, count rates are normalized
to count~s$^{-1}$\,half$^{-1}$,
ensuring consistency with the
count rates present in the ME database. 

Information on the 992 sources with single identifications
is presented in Table~3. The first column is a designation
in the form EXMS~BHHMM+DDd, where HHMM and DDd are the
RA and Decl. of the uncertainty region centroid in epoch
1950 coordinates. (Note: in order to maximise the usefulness of
Tables~3, 4 and 5, all coordinates therein are
presented for epoch 2000, although all
prior catalog searches and computations 
were performed using  epoch
1950 coordinates).

Most detections have a unique EXMS classification. 
In cases where the classification is
not unique, designators in the form (A), (B), etc are
appended.
In order to usefully group identifications,
detections have been ordered in
RA of the proposed candidate. Detections with
the same coordinate (but unique designator) 
may be split between tables depending on the
category of the identification.
 In cases 
where an
identified source was observed more than once, the entries
are additionally ordered by detection time.
The format for the time entries is YYYY/DDD HH:MM, indicating an
observation at MM minutes after HH hours on day DDD of year YYYY.
The proposed identifications are given (truncated to 18 characters),
followed by a ``type'' string which gives the first three letters
of the catalog from which the identification was drawn (see the list
of catalogs above), or 
the object category. 
The object categories are XRB: X-ray binaries;
SNR: supernova remnant; CLU: cluster of galaxies; and AGN: 
active galactic nucleus. 
 In this context we adopt the widest possible meaning of the term AGN,
  including Seyfert galaxies, BL Laceratae objects, 
  QSOs, quasars,  Radio Galaxies  and Optical Violent Variables (OVVs).
The next parameters are the collimator corrected count rate, 
together with the $\chi$$^{2}$ of the fit to the
light curve and the number of degrees of freedom (dof).
Large values of $\chi$$^{2}$/dof may indicate source variability.
Parameter $\lambda$ is 
the likelihood detection statistic which is the difference in
the logarithm of the likelihood between the best-fit and 
null-hypothesis models with no source present. 
In the null-hypothesis, $2\lambda$ is expected to be distributed
as $\chi$$^{2}$.
Flag values are given under the column labelled ``F'', 
(ranging between 2 and 4 for this sample).
Finally, the collimator distances, D, (CM in Fig.~\ref{fig:errorbox})
are given in arc minutes. Detections with collimator distances $\approxgt$30$'$
may contain large errors in their corrected count rates and
should be treated with caution. 
Remarks on individual entries are given in Sect.~\ref{subsect:remarks}.


The 58 detections with multiple candidates (Flag=1) are summarized in 
Table~4. The names of the
proposed candidates are stored in a string, the first
20 characters of which are printed in Table~4, together with the 
number of possible identifications, ${\rm N_{mult}}$.
The count rates given are raw values, uncorrected for collimator
transmission. More information on the proposed candidates
is given in Sect.~\ref{subsect:remarks}.

Table~5 contains information on the 160 unidentified detections 
(Flag=0). In this case, in addition to the coordinates of the
centroid of the uncertainty region,
positions of the four
bounding corners are also given (in decimal format for brevity), under columns
labelled C1 to C4. Again, the count rates are uncorrected for
collimator transmission.
  
\subsection{Remarks on individual catalog entries}
\label{subsect:remarks}

Comments are provided for individual catalog entries,
referenced by their EXMS designations (see Tables~3-5), below.
\smallskip

EXMS~B0016-722 --- AQ~Tuc lies within the uncertainty region and is
within 10$'$ of 5 X-ray sources. Globular cluster 47 Tuc is nearby.

EXMS~B0029-841 --- Four PKS sources lie within the uncertainty 
region, but none are cataloged X-ray emitters.

EXMS~B0110+649 ---  The XRB 2S\thinspace0114+65 is nearby.

EXMS~B0113-417 --- UV Phe lies within the uncertainty region, but is
not a cataloged X-ray emitter. The RASSBSC source adopted as the 
identification is consistent with the location of a cataloged X-ray emitting
Seyfert galaxy.

EXMS~B0114+005 ---  1RXSJ011704.2+000025 may be associated with 
E\thinspace0114.4$-$0015,
a Seyfert 1. 

EXMS~B0121+341 --- NGC 513 lies within the uncertainty region. It is
not a cataloged X-ray emitter but is bright and close.

EXMS~B0245-420 --- DK Eri lies within the uncertainty region, but
is not a cataloged X-ray emitter.

EXMS~B0250+415 --- Two X-ray counterparts of NGC 1129. Multiple
entries are 1H\thinspace0251+414 and IPC\thinspace025113+41.

EXMS~B0250+417 --- 1H\thinspace0251+414  and IPC\thinspace025113+41 
are two X-ray counterparts
of NGC 1129. 

EXMS~B0250+418 --- see above.

EXMS~B0252-417 --- PKS\thinspace0252$-$41 lies within the uncertainty region,
but is not a cataloged X-ray emitter.

EXMS~B0252+415 --- an unidentified source possibly associated with the
above object. Another unidentified source, EXMS~B0252+411, lies less
than one degree away.

EXMS~B0252+411 --- see above.

EXMS~B0309-317 --- PKS\thinspace0309$-$31 lies within the uncertainty region,
but is not a cataloged X-ray emitter.

EXMS~B0310+411 --- May be NGC 1275 or Mrk1073.

EXMS~B0402-654 --- PKS\thinspace0403$-$65 lies within the uncertainty region
but is not a cataloged X-ray emitter. 

EXMS~B0431-613B --- PKS\thinspace0429$-$61 lies within the uncertainty region,
but is not a cataloged X-ray emitter.

EXMS~B0439-129 --- PKS\thinspace0436$-$129 lies within the uncertainty region
but is not a cataloged X-ray emitter.

EXMS~B0446+447 --- 4C +44.12  and 4C +44.13, cataloged X-ray emitting
sources in 3C129, lie within the uncertainty region. 

EXMS~B0525-329 --- MS~05267$-$3301 lies within the uncertainty region,
but is not a cataloged X-ray emitter.

EXMS~B0529-650 --- The proposed RASSBSC identification is near the
transient X-ray binary EXO\thinspace53109$-$66.

EXMS~B0559-664 --- possibility of confusion with A\thinspace0535$-$668. 
Four RASSBSC sources within uncertainty region.

EXMS~B0532-664B --- see above.

EXMS~B0533-663B --- see above.

EXMS~B0533-662A --- see above.

EXMS~B0534-657 --- see above.

EXMS~B0537-713 --- Probably an LMC object; LMC X-1 and X-2 were both
slewed over near the detection time, although neither fell within
the uncertainty region, but the raw (and collimator corrected) 
count rate is too high for either. PKS\thinspace0531$-$71 lies within the  
uncertainty region, but is not a cataloged X-ray emitter. 

EXMS~B0601-701 --- an unidentified source less than one degree from
another unidentified source, EXMS~B0606-697.

EXMS~B0606-697 --- see above.

EXMS~B0607-712 --- PKS\thinspace0611$-$71 lies within the uncertainty region, 
but is not a cataloged X-ray emitter. An unidentified source, EXMS~B0610-714,
lies less than one degree away.

EXMS~B0610-714 --- see above.

EXMS~B0613+228 --- Two obscure radio sources, PKS\thinspace0615+22 and 
4C +22.14, neither cataloged as an X-ray emitter.

EXMS~B0613+229 --- 4C +22.14 lies within the uncertainty region,
but is not a cataloged X-ray emitter.

EXMS~B0629-713 --- an unidentified source less than one degree from
another unidentified source, EXMS~B0640-715.

EXMS~B0640-715 --- see above.

EXMS~B0648-698 --- PKS\thinspace0650$-$70 lies within the uncertainty region, 
but is not a cataloged X-ray emitter.

EXMS~B0654-559 --- PKS\thinspace0649$-$55 lies within the uncertainty region,
but is not a cataloged X-ray emitter.

EXMS~B0655-707 --- an unidentified source less than one degree from another
unidentified source, EXMS~B0656-697.

EXMS~B0656-697 --- see above.

EXMS~B0658+753 --- 4C +74.12 lies within the uncertainty region, but
is not a cataloged X-ray emitter. 

EXMS~B0821-425A --- 1ES\thinspace0821$-$426 is a cataloged X-ray source.

EXMS~B0821-425B --- an unidentified source which might be associated
with the above, although it does not lie within the uncertainty region.

EXMS~B0820-424 --- see above.

EXMS~B0823-426 --- see above.

EXMS~B0834-428 --- probably G0834-430 and/or 4U0836-429; both were
slewed over near the detection time though neither falls within the
uncertainty region.

EXMS~B0834+254 --- Two extragalactic objects, Mrk1218 and 
B2\thinspace0834+25 lie within the uncertainty region but only Mrk1218 is
a cataloged X-ray emitter. 

EXMS~B0912+354 --- 4C +34.31 lies within the  uncertainty region,
but is not a cataloged X-ray emitter.

EXMS~B0917-549 --- PKS\thinspace0916$-$54 lies within the uncertainty region, but
is not a cataloged X-ray emitter.

EXMS~B0923-308 --- PKS\thinspace0923$-$30 lies within the uncertainty region,
but is not a cataloged X-ray emitter. The proposed RASSBSC identification
is near X-ray source 1H\thinspace0919$-$312.

EXMS~B1042-599 --- Eta Carinae lies within the uncertainty region, but
is inconsistent with the position of the RASSBSC source.

EXMS~B1044-595 --- see above.

EXMS~B1043-593 --- see above.

EXMS~B1040-593 --- see above.

EXMS~B1049+385 --- B2\thinspace1049+38, a high-redshift radio galaxy
(or possibly Seyfert 2), lies within the uncertainty region but is
not a cataloged X-ray emitter. 

EXMS~B1123-588 --- The proposed RASSBSC identification may be
associated with the X-ray emitting SNR MSH11-54.

EXMS~B1153+317 --- 4C +31.38 lies within the uncertainty region
but is not a cataloged X-ray emitter.

EXMS~B1213+038 --- The uncertainty region contains several radio
sources, one of which (4C +04.41) corresponds to X-ray source 
1ES\thinspace1215+039.

EXMS~B1155-187 --- Three obscure emission line galaxies lie within
the uncertainty region.

EXMS~B1235+708 --- 4C +70.13 lies within the  uncertainty region,
but is not a cataloged X-ray emitter.

EXMS~B1246-410B ---  ESO 323-G32 is a Seyfert 2 in cluster A3526.

EXMS~B1254+276 --- multiple unclassified AGN in Veron catalog. 
Multiple entries all have name UNKNOWN.

EXMS~B1324-312 ---  A3558 is a cataloged X-ray emitter. 

EXMS~B1415+253 ---  NGC 5548 and 1E\thinspace14156+259 are two X-ray emitting
AGN. 

EXMS~B1415+255 --- see above.

EXMS~B1416+256 --- see above.

EXMS~B1517-613 --- TrA X-1 was slewed over near the detection time but
would have appeared near the edge of the wrong quadrant.

EXMS~B1525+525 --- 4C +52.35 lies within the uncertainty region
but is not a cataloged X-ray emitter.

EXMS~B1550-609 --- 4U1543-624 and 4U1556-605 were both slewed over near
the detection time, but neither fell within the uncertainty region.

EXMS~B1611-508 --- HW Nor is within 10$'$ of 18 X-ray sources.

EXMS B1637-671 --- 4U1627-673 was slewed over near the time of detection
but would have appeared in the wrong quadrant.

EXMS~B1658-228A --- an unidentified source less than a degree from
another unidentified source, EXMS~B1658-228B.

EXMS~B1658-228B --- see above.

EXMS~B1702-429 --- probably 4U1705-440 and/or 4U1702-429; both were 
slewed over near the detection time though neither falls within the 
uncertainty region.

EXMS~B1719-436 --- an unidentified source less than a degree from
another unidentified source, EXMS~B1722-442.

EXMS~B1722-442 --- see above.

EXMS~B1721-231A --- an unidentified source less than a degree from
two other unidentified sources, EXMS~B1721-231B and EXMS~B1721-231C.

EXMS~B1721-231B --- see above.

EXMS~B1721-231C --- see above.

EXMS~B1730-445 --- 4U1735-444 passed slightly beyond the field of
view near the time of the detection.

EXMS~B1734-155 --- TX Ser lies within the uncertainty region and
is within 10$'$ of 1RXS{\thinspace}J173735.1$-$152357 
and 2E\thinspace1734.7$-$1522.

EXMS~B1741-337 --- PKS\thinspace1742$-$337 lies within the uncertainty region,
but is not a cataloged X-ray emitter.

EXMS~B1743-300A --- probably GC X-1 or one of the many nearby objects,
most of which were slewed over but none of which fell within the
uncertainty region.

EXMS~B1743-300B --- see above.

EXMS~B1743-303B --- probably one or more of the many Galactic Center objects
including SL1744-300 and neighboring sources, many of which were slewed
over but none of which fell within the uncertainty region.

EXMS~B1747-366 --- probably 4U1746-360 and/or A1744-361; both were slewed
over near the detection time though neither falls within the uncertainty
region.

EXMS~B1754-289 --- an unidentified source less than a degree from another
unidentified source, EXMS~B1755-295.

EXMS~B1755-295 --- see above.

EXMS~B1827-100 --- an unidentified source less than a degree from another
unidentified source, EXMS~B1828-100.

EXMS~B1828-100 --- see above.

EXMS~B1837-052 --- Near the SNR G27.4+0.0.

EXMS~B1846-029A --- In addition to EXO1846-03, G1845-03 and A1845-024 were 
also slewed over near the detection time but neither fell within the 
uncertainty region.

EXMS~B1905+009 --- Probably 4U1905+000 and/or Aql X-1, both of which were
slewed over but neither of which fell within the uncertainty region. 
V810 Aql does lie within the uncertainty region and
within 10$'$ of five Rosat sources. 

EXMS~B1906+007 --- 4C +00.71 lies within the uncertainty region, but
is not a cataloged X-ray emitter.

EXMS~B1931+800 --- 4C +79.20 lies within the uncertainty region,
but is not a cataloged X-ray emitter.

EXMS~B1948+113 --- 4C +11.59 lies within the uncertainty region,
but is not a cataloged X-ray emitter.

EXMS~B1955+123 --- an unidentified source less than a degree
from 4U1957+115, and also another unidentified source, EXMS~B1955+124.

EXMS~B1955+124 --- see above.

EXMS~B2124+751 --- 4C +74.27 lies within uncertainty region, but is
not a cataloged X-ray emitter.

EXMS~B2216-389 ---  Q2217$-$391, an extragalactic object, lies within
the uncertainty region, but is not a cataloged X-ray emitter. 1H\thinspace2217$-$392 is a nearby stellar X-ray source.

EXMS~B2232-734 --- PKS\thinspace2238-73 lies within the uncertainty region 
but is not a cataloged X-ray emitter.

EXMS~B2233-657 --- PKS\thinspace2239-65 lies within the  uncertainty region, 
but is not a cataloged X-ray emitter.

EXMS~B2223-331 --- PKS\thinspace2224$-$33 lies within the uncertainty region,
but is not a cataloged X-ray emitter.

\subsection{Systematic effects and accuracy of derived parameters}
\label{subsect:systematics}

In order to establish that the catalog is free of
obvious systematic errors, a large subsample of
sources was studied in detail. The adopted subsample
contains 463 slew sources which were assigned identifications from the 
XRBCAT database of X-ray binaries (van Paradijs 1995), and where the
identifications lie within the formal uncertainty regions
(Flag=4).
Apart from the
fact that this is the most numerous category of counterpart in the
EXMS, these X-ray binaries 
have well defined positions
and tend to be bright X-ray sources.
This helps ensure that they
can be detected near the edge of the FOV.
The disadvantage with this
population is their strong concentration towards the galactic plane, 
and their intrinsic variability.
The behavior of the offsets between these sources and the center
of their uncertainty regions was simulated and
studied, as is now discussed.

\begin{figure}
\begin{center}
\epsfig{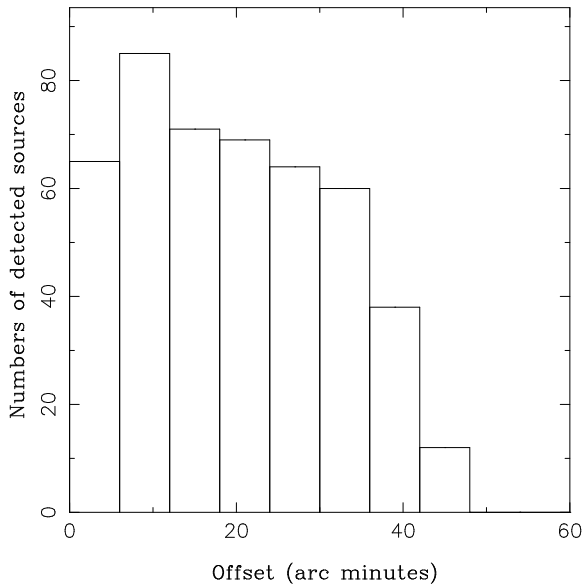}
\caption{The histogram of observed source offsets (in 6\farcm0
 bins) for the 463 sources identified as X-ray binaries whose positions
are within the formal uncertainty regions}
\label{fig:real_histogram}
\end{center}
\end{figure}

\begin{figure}
\begin{center}
\epsfig{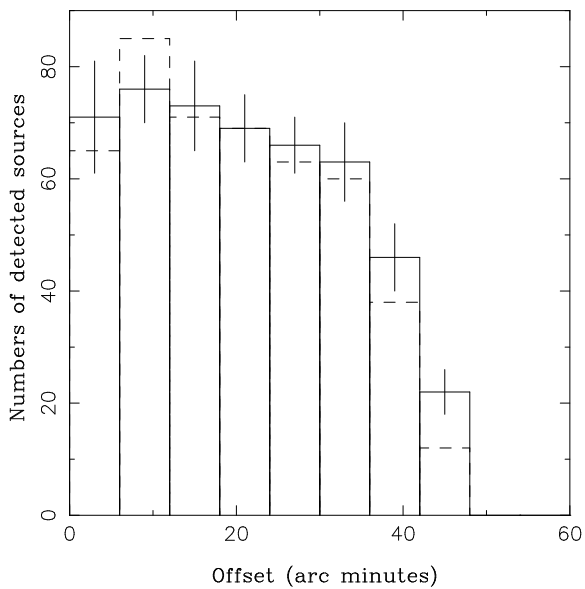}
\caption{A simulated histogram of source offsets (in 6\farcm0
 bins) for 500 sources, produced by averaging ten simulations. 
The variance in
each bin is shown by a vertical line. The observed histogram
(Fig.~\ref{fig:real_histogram}) is plotted using a broken line}
\label{fig:simulated_histogram}
\end{center}
\end{figure}

\subsubsection{FOV offset simulation}

The offset of a given detection in the EXMS is simply the angular
distance between the centroid of the uncertainty region and the
catalog coordinates of the proposed identification. Since it is
required that all identifications passed through the
FOV, the maximum possible
offset is given by the 45$'$ FWHM of the collimator response.
For simplicity the offsets are broken down into two 
components, defined in Fig.~\ref{fig:errorbox}. The absolute
offset for source E is the scalar
distance EC and the vector
offset is EM. Similar
offsets are defined for internal sources.

Absolute offsets were derived for the 463 XRBCAT identifications in the
EXMS, and
these were then summed into bins of 6\farcm0, resulting in the
histogram of Fig.~\ref{fig:real_histogram}. 
In order to understand the shape of
this distribution, a simulation was performed in the following
manner. First, a population of X-ray binaries was created with
a realistic count rate distribution. This was achieved by correlating the
XRBCAT against the ME database to obtain normalised ME counts for
each matching source. Sources were then drawn randomly from this
population and assigned random positions along a line, 
approximating the narrow uncertainty regions of the true sample.
The {\it observed}\, counts of each source were then adjusted
in accordance with the distance of the source from the center of
the line, using the ME collimator response fuction. The average
background count rate of the 1--8~keV lightcurves
is $\sim$\,15 counts\,s$^{-1}$\,half$^{-1}$,
and this was adopted in the simulations. Since the background
events are predominantly particle-induced, they are independent
of collimator distance.
As a consequence, faint sources
are not detected unless they fall near the middle of the
simulated FOV, but as such sources are more numerous, they play an
important role in the shape of the offset histogram. 
The procedure was repeated
until $\approx$460 sources had
been detected. Since there are $<$100 detections
in each
bin, the detailed shape of the histogram is dependent on
statistical fluctuations. In order to characterize the shape and
variance of a mean 
histogram, the results of ten independent simulations were averaged.
The overall shape of the observed histogram is reproduced satisfactorily,
with an approximately flat-topped distribution out to about 0.6 of the 
maximum offset (Fig.~\ref{fig:simulated_histogram}). 
In most bins, the real and 
simulated histograms agree
within the variance of the simulation, and the small discrepancies that
exist can be understood in terms of limitations in the simulation,
particularly in the manner in which the detection threshold is modeled,
and the fact that a few bright sources contribute predominantly to
the detected numbers of sources in the high-offset bins.

\begin{figure}
\begin{center}
\epsfig{file=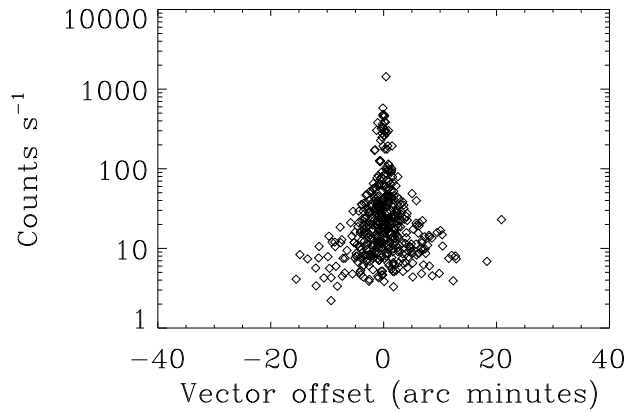, width=3.5in, angle=0.0}
\caption{Distribution of vector offsets (component of offset
perpendicular to the direction of slew; vector EM in
Fig.~\ref{fig:errorbox}) for the 463 sources indentified
with X-ray binaries which also lie within the formal
uncertainty regions}
\label{fig:vectors_463}
\end{center}
\end{figure}

Having established that the source offsets for this subsample are
free of obvious systematic effects,
the behavior of the vector component of the offsets which lie in the
slew direction was examined. 
In the absence of systematic errors in the pointing
and timing information, the resultant distribution of offsets should
be clustered symmetrically around zero. 
This is indeed the case,
as shown in Fig.~\ref{fig:vectors_463}, where the 463 vector offsets 
are plotted against raw count rate. The mean vector offset is
consistent with zero: $(-0.01 \pm 0.20)$$'$. As is to be expected, the spread
in the offsets is largest for the faintest sources.

\begin{figure}
\begin{center}
\epsfig{file=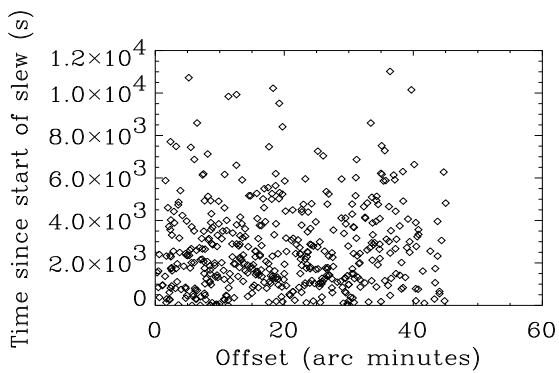, height=2.0in,width=3.0in, angle=0.0}
\caption{Plot of absolute offsets (the distance EC in 
Fig.~\ref{fig:errorbox})
against time since start of
first slew leg, for the 463 sources indentified
with X-ray binaries which also lie within their own error boxes. No
time-dependent effects are evident}
\label{fig:absolute_offsets}
\end{center}
\end{figure}

\begin{figure}
\begin{center}
\epsfig{file=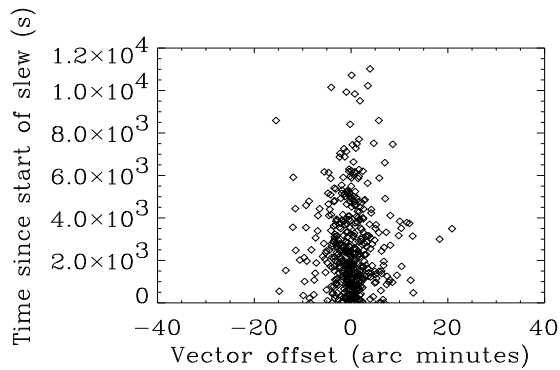, height=2.0in,width=3.0in, angle=0.0}
\caption{Plot of vector offsets against time since start of
first slew leg, for the same sample as in Fig.~\ref{fig:vectors_463}. 
Again no time-dependent effects are evident}
\label{fig:vectors_slew1}
\end{center}
\end{figure}

\begin{figure}
\begin{center}
\epsfig{file=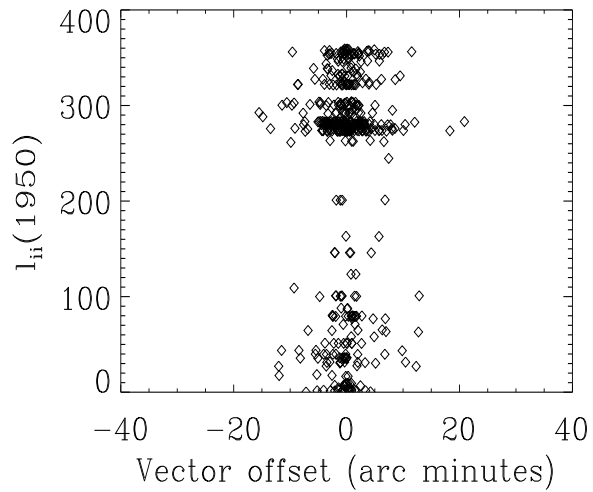, height=3.0in, width=3.5in, angle=0.0}
\caption{Plot of vector offsets against galactic longitude, l$_{II}$,
demonstrating the lack of any dependence}
\label{fig:vectors_gal_long}
\end{center}
\end{figure}

\begin{figure}
\begin{center}
\epsfig{file=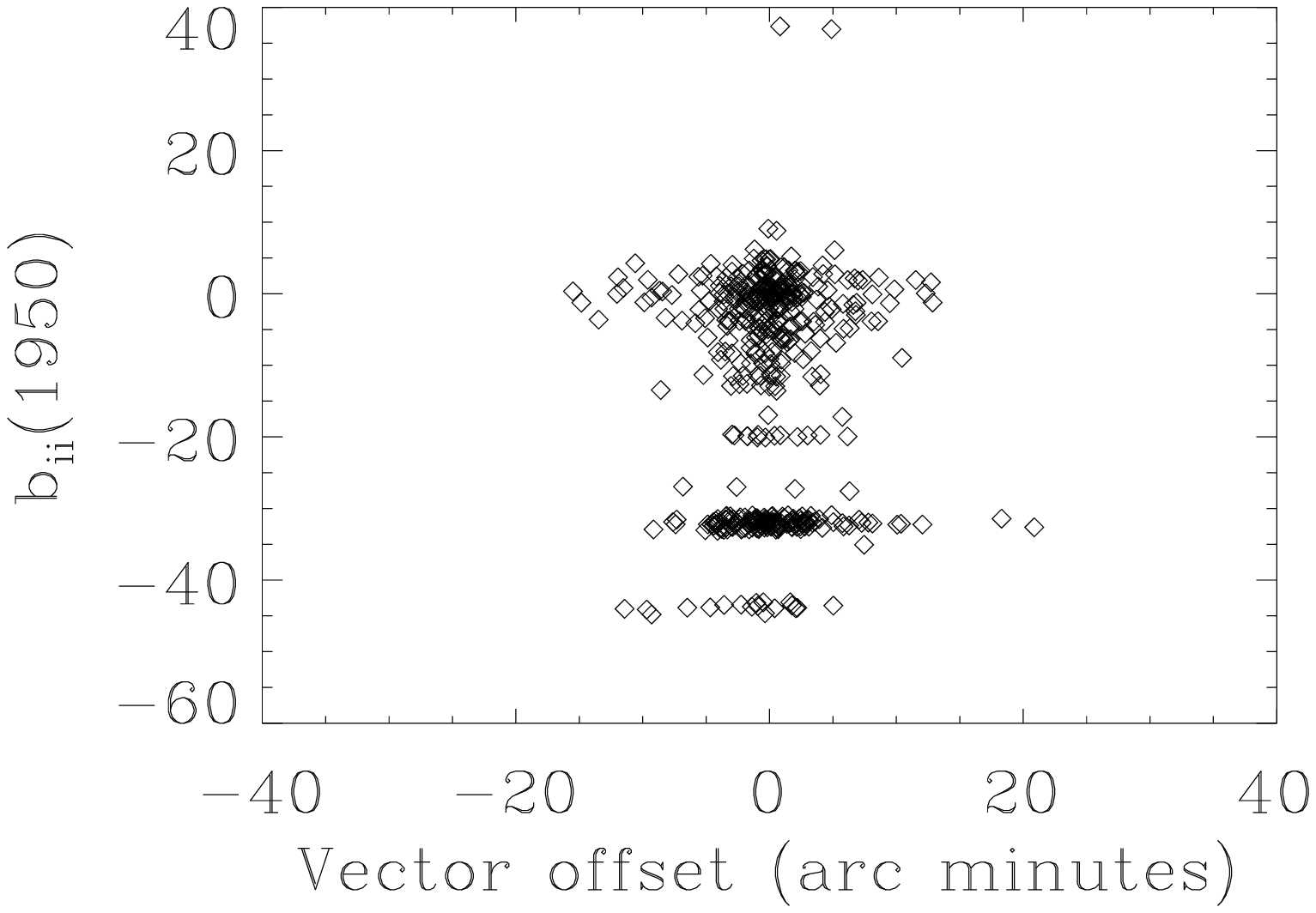, height=3.0in, width=3.5in, angle=0.0}
\caption{As for Fig.~\ref{fig:vectors_gal_long}, but for galactic 
latitude}
\label{fig:vectors_gal_lat}
\end{center}
\end{figure}

The behavior of the absolute and vector 
offsets as
a function of the elapsed time since the start of each slew
was also examined, to
ascertain whether the spacecraft attitude reconstruction gradually becomes
less accurate. Since the slew durations vary from
hundreds to thousands of seconds, the data span a wide time
range, but it is clear (see Figs.~\ref{fig:absolute_offsets} and 
\ref{fig:vectors_slew1}) that the offsets
appear random and independent of elapsed time. 

Finally, the distributions of the absolute and vector
offsets as a function of position in the sky were examined. 
Due to the sample being
drawn solely from X-ray binaries, the sources are clustered around the
Galactic plane and the Magellanic clouds. This means that there are large
areas of the sky where the offsets are not sampled. 
However, at least for the subsample, there is
no suggestion of any position-dependent offset effects in either 
equatorial or Galactic coordinates (see Figs.~\ref{fig:vectors_gal_long} and 
\ref{fig:vectors_gal_lat} for the
latter).

\subsubsection{Count rate accuracy}

Due to the variability of most X-ray sources, 
there are few sources in the EXMS which enable the accuracy
of the derived count rates to be estimated. The brightest sources, such
as X-ray binaries, are highly variable, while those sources which are 
time-invariant 
tend to be faint and extended (the latter leading to 
incorrectly estimated collimator corrections).
Nevertheless, the observations of
SNRs within the EXMS are in good agreement with 
the expected count rates based on ME pointed observations.

\begin{figure}
\begin{center}
\epsfig{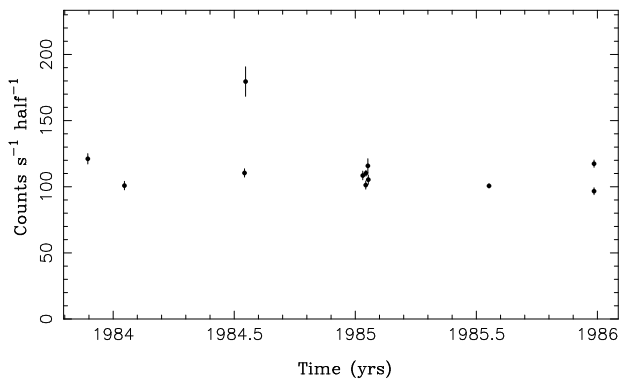}
\caption{EXMS collimator corrected count rates for the Cas-A SNR
(assuming a point source). The expected count rate is 108 
count\,s$^{-1}$\,half$^{-1}$. With the exception of the measurement
on 1984/201, which occured at an offset of 35$'$,
the count rates lie
within $\sim$ $15$\% of the expected value}
\label{fig:casa}
\end{center}
\end{figure}

\begin{figure}
\begin{center}
\epsfig{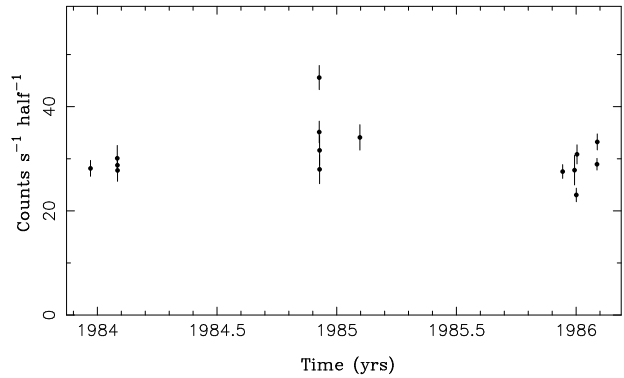}
\caption{Collimator corrected count rates for the Tycho SNR
(assuming a point source). The expected count rate is 32 
count\,s$^{-1}$\,half$^{-1}$, consistent with the mean
EXMS value of $31 \pm 5$ count\,s$^{-1}$\,half$^{-1}$}
\label{fig:tycho}
\end{center}
\end{figure}

As an example, in 12 detections of
Cas-A between 1983 and 1986, the derived EXMS count rates
vary between 97 and 180 count\,s$^{-1}$\,half$^{-1}$. Most
values are between 100 and 120 count\,s$^{-1}$\,half$^{-1}$, 
with a mean
value of 114 $\pm$ 9 count\,s$^{-1}$\,half$^{-1}$ 
(see Fig.~\ref{fig:casa}). The detection with the highest 
count rate is also that with the largest offset distance of 35$'$ (see
Table~3). The extended nature of the source leads to
an overestimated collimator correction. However, the mean value
is in good agreement with the count rate during the single
pointed ME observation of Cas-A of 108 count\,s$^{-1}$\,half$^{-1}$.
Similar agreement is seen for the 15 slew detections of
Tycho, with a mean value of 31 $\pm$ 5
count\,s$^{-1}$\,half$^{-1}$, compared with 32 count\,s$^{-1}$\,half$^{-1}$
from two pointed observations
(Fig.~\ref{fig:tycho}).

\subsubsection{Count rate to flux conversion}

\begin{figure}
\begin{center}
\epsfig{file=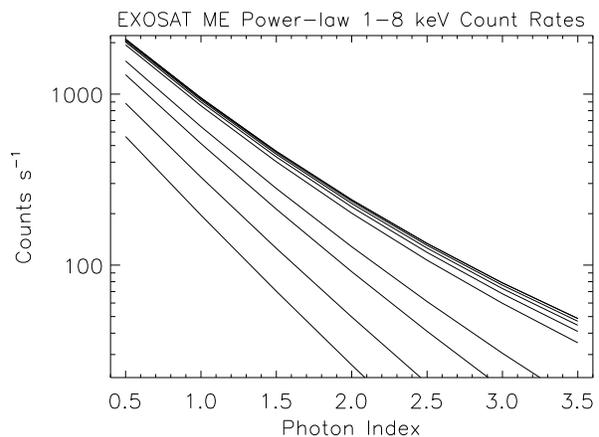, width=3.2in, angle=0}
\caption{Curves for converting from EXOSAT ME count rates to
deadtime corrected fluxes for power-law spectra.  
From top to bottom, the curves give the predicted count rates
for a spectral normalization of unity for
N${\rm _H}$ values
of $<$$10^{21}$, $10^{21}$, $2.5 \times 10^{21}$,
$5 \times 10^{21}$, $10^{22}$, $5 \times 10^{22}$,
$10^{23}$, $2.5 \times 10^{23}$, and $5 \times 10^{23}$~\hcm}
\label{fig:me_efficiency}
\end{center}
\end{figure}

An approximate conversion from count rate to flux may be obtained
by noting that 
1 ME count\,s$^{-1}$\,half$^{-1}$ in the energy range 2--10~keV is
approximately equal to $1.2 \times 10^{-11}$~erg~cm$^{-2}$~s$^{-1}$.
This conversion is dependent on spectral shape and
curves for converting from the EXOSAT ME count rates given in 
Tables~3-5 to 1--8~keV fluxes are given in Fig.~\ref{fig:me_efficiency}. 
This figure gives the conversion factors
for a power-law model with photon indices between 0.5 and 3.5
and absorptions $<$$5 \times 10^{23}$~\hcm.
The {\it cataloged}\, 
1--8~keV count rate, $c$, should first be corrected for deadtime losses
by multiplying it by an approximate deadtime factor, $d{\rm _{time}} 
 = 1.09 + 2.56\times 10^{-4}~c + 5.6 
\times 10^{-8}~c^{2}$.
Comparison with the count rate read from Fig.~\ref{fig:me_efficiency} 
gives the
power-law spectral normalization at 1~keV, and hence the flux.


 
\section{Summary}
\label{sect:summary}

The EXMS, a new catalog of X-ray detections 
derived from observations made by EXOSAT during slew manoeuvres between
1983 and 1986, is presented. 
Where possible,
detections are identified with cataloged sources and the raw count rates
corrected for collimator losses. Many types of object are
detected, although X-ray binaries constitute the most common 
identification. The catalog has been shown to be free of obvious
systematic errors and to contain reliable count rates
for time-invariant sources. 
An electronic version of this catalog is maintained within the EXOSAT
database and archive system at ESTEC
(telnet://xray@exosat.estec.esa.nl).

\begin{figure}
\begin{center}
\epsfig{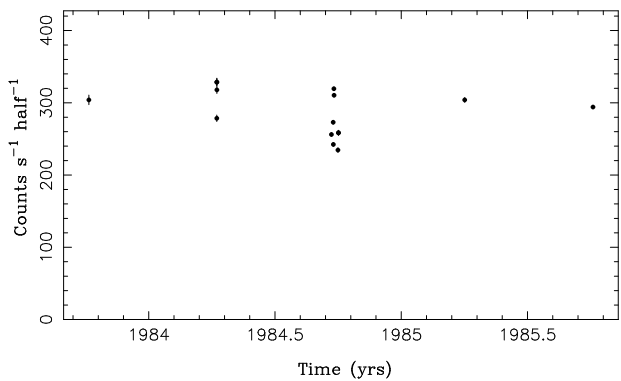}
\caption{EXMS detections of the XRB Ser X-1.
Two pointed observations performed on 1985/251
and 252 recorded count rates of 344 and 343 count\,s$^{-1}$\,half$^{-1}$,
similar to the EXMS values}
\label{fig:serx1}
\end{center}
\end{figure}

\begin{figure}
\begin{center}
\epsfig{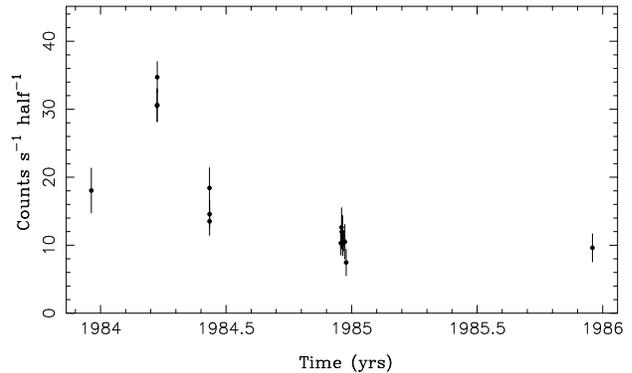}
\caption{EXMS detections of the XRB 4U\thinspace0918$-$549 showing
an apparant long-term decrease in 1--8~keV intensity}
\label{fig:4u0918}
\end{center}
\end{figure}

\begin{figure}
\begin{center}
\epsfig{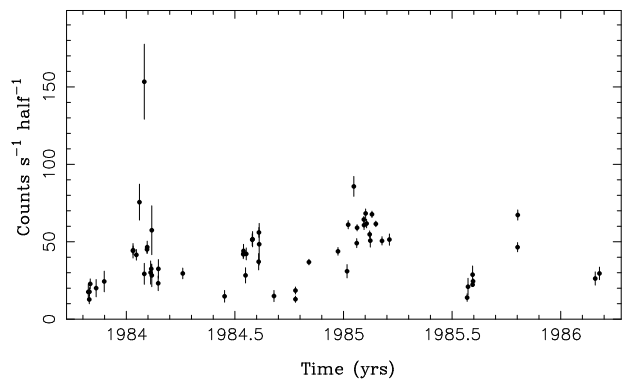}
\caption{EXMS detections of the XRB LMC X-3. The 198 (or possibly 99) day
periodicity discovered by Cowley et al (1991), may be evident}
\label{fig:lmcx3}
\end{center}
\end{figure}

The catalog contains new data on the long-term time variability
of many well known sources and in particular on X-ray binaries. 95
different X-ray binaries appear in Table~3; roughly half of all 
X-ray binaries which have ever been detected. 
All of the expected
X-ray binaries were seen, many of them repeatedly. We note that the
transient source EXO\,0748-676 appears as an EXMS source
several times before it was actually detected in February 1985.
Repeated
source detections tend to be grouped at six monthly intervals due to
the way in which EXOSAT slewed (see Sect.~1).
For instance, Ser~X-1 
(Fig.~\ref{fig:serx1}) was
detected on 15 occasions between 1983/279 and 1985/278, on
nine separate days in five clusters each spaced by six months.
Lightcurves for the XRBs 4U\thinspace0918$-$549 and LMC~X-3 
are presented in Figs.~\ref{fig:4u0918}--\ref{fig:lmcx3},
together with brief comments.

\begin{acknowledgements} 
We thank H. Siddiqui, A. Hazell, K. Bennett and E. Kuulkers.
F. Ochsenbein of the Observatoire Astronomique de 
Strasbourg is thanked for help on the EXMS naming policy.
In addition to the online databases at ESTEC, 
this research made use of data obtained from the High Energy
Astrophysics Science Archive Research Center, provided by the
NASA-Goddard Space Flight Center. We thank the referee, Dr W. Voges,
for many helpful suggestions.
\end{acknowledgements}

\vfill
\eject

\thispagestyle{empty}
\begin{landscape}

\thispagestyle{empty}
\begin{center}
\begin{tabular}{llllllcrlll}
\multicolumn{10}{l}{{\bf {Table 3.}} All singly-identified EXMS entries}\\
\hline
Designation & RA & Dec & Time & Identification
& Type & Counts & ${\chi}^{2}$/dof & $\lambda$ & F & D \\
            &  (2000) & (2000) &      &              
&       & (s$^{-1}$\,half$^{-1}$) &   &  &  & (arc min) \\
\hline
EXMS B0022+638  & 00 25 22 & +64 07 26 & 1983/355 14:10 &              Tycho &  SNR &    28.2\,$\pm$  1.5 &    0.98/ 4 &   169.6 & 3 &  1.3 \\
EXMS B0021+638B & 00 24 45 & +64 05 26 & 1984/031 18:01 &              Tycho &  SNR &    28.8\,$\pm$  1.2 &   33.57/12 &   293.5 & 3 &  4.7 \\
EXMS B0021+638C & 00 24 30 & +64 06 11 & 1984/031 21:48 &              Tycho &  SNR &    27.8\,$\pm$  2.1 &    4.08/ 5 &   111.1 & 3 &  6.2 \\
EXMS B0021+638A & 00 24 05 & +64 05 55 & 1984/031 12:49 &              Tycho &  SNR &    30.1\,$\pm$  2.5 &   18.76/ 5 &    75.2 & 3 &  8.1 \\
EXMS B0022+641A & 00 25 23 & +64 23 45 & 1984/340 05:14 &              Tycho &  SNR &    45.6\,$\pm$  2.3 &    6.99/11 &   195.8 & 3 & 15.2 \\
\hline
\end{tabular}
\end{center}
\end{landscape}
\newpage
\thispagestyle{empty}
\begin{landscape}
\setcounter{table}{3}
\begin{center}
\begin{tabular}{llllllcrlll}
\multicolumn{10}{l}{{\bf {Table 3}} (continued).}\\
\hline
Designation & RA & Dec & Time & Identification
& Type & Counts & ${\chi}^{2}$/dof & $\lambda$ & F & D \\
            & (2000) & (2000) &      &              
&       & (s$^{-1}$\,half$^{-1}$) &  &  &  & (arc min) \\
\hline
EXMS B0107-735  & 01 09 07 & -73 14 47 & 1984/299 08:11 &       SMC X-1      &  XRB &    43.4\,$\pm$  5.3 &    6.32/11 &    33.5 & 4 & 36.0 \\
EXMS B0107-734  & 01 09 02 & -73 12 50 & 1984/299 08:08 &       SMC X-1      &  XRB &    39.1\,$\pm$  5.2 &   10.69/11 &    28.4 & 4 & 36.8 \\
EXMS B0901-402A & 01 17 45 & -73 16 23 & 1984/312 15:04 &      SMC X-1       &  XRB &    26.4\,$\pm$  4.9 &    0.29/ 2 &    14.6 & 3 & 10.5 \\
EXMS B0113-735  & 01 15 09 & -73 18 50 & 1985/124 06:33 &      SMC X-1       &  XRB &    23.1\,$\pm$  1.6 &    6.10/ 5 &   107.9 & 3 &  5.2 \\
EXMS B0115-732  & 01 16 46 & -72 59 14 & 1985/125 02:07 &       SMC X-1      &  XRB &    20.6\,$\pm$  3.6 &    1.21/ 4 &    16.3 & 4 & 25.3 \\
\hline
\end{tabular}
\end{center}
\end{landscape}
\newpage
\thispagestyle{empty}
\begin{landscape}
\setcounter{table}{3}
\begin{center}
\begin{tabular}{llllllcrlll}
\multicolumn{10}{l}{{\bf {Table 3}} (continued).}\\
\hline
Designation & RA & Dec & Time & Identification
& Type & Counts & ${\chi}^{2}$/dof & $\lambda$ & F & D \\
            & (2000) & (2000) &      &              
&       & (s$^{-1}$\,half$^{-1}$) &  &  &  & (arc min) \\
\hline
EXMS B0431-613B & 04 32 14 & -61 17 25 & 1984/285 23:59 &   J043126.6-612622 &  RAS &     6.5\,$\pm$  1.0 &    6.50/11 &    21.6 & 4 & 10.6 \\
EXMS B0446+449  & 04 50 29 & +45 00 23 & 1984/066 20:22 &   J045017.2+450217 &  RAS &     8.0\,$\pm$  1.8 &    0.10/ 1 &     9.6 & 4 &  1.1 \\
EXMS B0447-133  & 04 50 17 & -13 15 13 & 1983/262 20:35 &   J044954.6-130544 &  RAS &     5.8\,$\pm$  1.6 &   14.09/ 5 &     7.4 & 4 &  9.2 \\
EXMS B0452+047  & 04 54 58 & +04 46 59 & 1984/286 01:07 &   J045554.7+044023 &  RAS &    19.8\,$\pm$  7.3 &    0.00/ 5 &     9.7 & 4 &  4.5 \\
EXMS B0512-402  & 05 14 01 & -40 10 28 & 1985/057 21:08 &       4U0512-401   &  XRB &     8.7\,$\pm$  1.4 &    6.25/11 &    19.4 & 4 &  2.6 \\
\hline
\end{tabular}
\end{center}
\end{landscape}
\newpage
\thispagestyle{empty}
\begin{landscape}
\setcounter{table}{3}
\begin{center}
\begin{tabular}{llllllcrlll}
\multicolumn{10}{l}{{\bf {Table 3}} (continued).}\\
\hline
Designation & RA & Dec & Time & Identification
& Type & Counts & ${\chi}^{2}$/dof & $\lambda$ & F & D \\
            & (2000) & (2000) &      &              
&       & (s$^{-1}$\,half$^{-1}$) &  &  &  & (arc min) \\
\hline
EXMS B0518-718C & 05 17 53 & -71 46 37 & 1984/291 22:02 &       LMC X-2      &  XRB &    28.4\,$\pm$  2.1 &    4.33/ 5 &   102.2 & 4 & 17.1 \\
EXMS B0523-721C & 05 22 11 & -72 06 39 & 1984/293 16:00 &       LMC X-2      &  XRB &    27.2\,$\pm$  2.7 &    1.58/ 1 &    56.1 & 4 &  9.2 \\
EXMS B0520-719B & 05 19 19 & -71 54 18 & 1984/294 23:57 &       LMC X-2      &  XRB &    30.7\,$\pm$  1.1 &    7.82/11 &   392.5 & 4 &  4.2 \\
EXMS B0524-722  & 05 24 06 & -72 12 08 & 1984/298 12:42 &      LMC X-2       &  XRB &    36.5\,$\pm$  4.5 &    0.91/ 5 &    38.6 & 3 & 22.1 \\
EXMS B0519-718B & 05 18 17 & -71 49 23 & 1984/299 07:48 &       LMC X-2      &  XRB &    29.1\,$\pm$  2.0 &    2.08/ 4 &   108.9 & 4 & 14.0 \\
\hline
\end{tabular}
\end{center}
\end{landscape}
\newpage
\thispagestyle{empty}
\begin{landscape}
\setcounter{table}{3}
\begin{center}
\begin{tabular}{llllllcrlll}
\multicolumn{10}{l}{{\bf {Table 3}} (continued).}\\
\hline
Designation & RA & Dec & Time & Identification
& Type & Counts & ${\chi}^{2}$/dof & $\lambda$ & F & D \\
            & (2000) & (2000) &      &              
&       & (s$^{-1}$\,half$^{-1}$) &  &  &  & (arc min) \\
\hline
EXMS B0538-642B & 05 38 23 & -64 13 54 & 1984/012 09:55 &       LMC X-3      &  XRB &    44.5\,$\pm$  2.7 &    2.03/ 5 &   132.4 & 4 &  9.1 \\
EXMS B0538-641A & 05 39 09 & -64 08 10 & 1984/017 21:46 &       LMC X-3      &  XRB &    41.5\,$\pm$  3.4 &    0.34/ 1 &    72.8 & 4 &  0.6 \\
EXMS B0537-642A & 05 37 22 & -64 11 18 & 1984/023 02:46 &       LMC X-3      &  XRB &    75.6\,$\pm$ 11.6 &    0.89/ 2 &    21.8 & 4 &  9.7 \\
EXMS B0543-635  & 05 43 25 & -63 31 10 & 1984/031 06:44 &       LMC X-3      &  XRB &   153.3\,$\pm$ 24.2 &    1.73/ 4 &    20.2 & 4 & 38.0 \\
EXMS B0534-645A & 05 35 10 & -64 29 19 & 1984/031 06:43 &       LMC X-3      &  XRB &    29.3\,$\pm$  6.8 &    0.92/ 5 &     9.3 & 4 & 34.1 \\
\hline
\end{tabular}
\end{center}
\end{landscape}
\newpage
\thispagestyle{empty}
\begin{landscape}
\setcounter{table}{3}
\begin{center}
\begin{tabular}{llllllcrlll}
\multicolumn{10}{l}{{\bf {Table 3}} (continued).}\\
\hline
Designation & RA & Dec & Time & Identification
& Type & Counts & ${\chi}^{2}$/dof & $\lambda$ & F & D \\
            & (2000) & (2000) &      &              
&       & (s$^{-1}$\,half$^{-1}$) &  &  &  & (arc min) \\
\hline
EXMS B0540-644  & 05 40 39 & -64 23 09 & 1985/038 06:00 &       LMC X-3      &  XRB &    68.3\,$\pm$  2.9 &    7.11/11 &   282.9 & 4 & 21.1 \\
EXMS B0541-638  & 05 41 42 & -63 51 40 & 1985/039 20:41 &       LMC X-3      &  XRB &    61.8\,$\pm$  2.8 &    3.63/ 4 &   240.5 & 4 & 22.3 \\
EXMS B0540-639C & 05 40 53 & -63 56 12 & 1985/044 22:05 &       LMC X-3      &  XRB &    54.8\,$\pm$  2.3 &    5.70/ 5 &   289.2 & 4 &  0.6 \\
EXMS B0535-643D & 05 35 22 & -64 16 58 & 1985/045 23:50 &       LMC X-3      &  XRB &    50.8\,$\pm$  4.2 &    1.05/ 5 &    72.1 & 4 &  0.6 \\
EXMS B0539-641B & 05 39 58 & -64 05 03 & 1985/048 19:16 &      LMC X-3       &  XRB &    67.7\,$\pm$  2.1 &    2.42/ 5 &   540.2 & 3 &  6.2 \\
\hline
\end{tabular}
\end{center}
\end{landscape}
\newpage
\thispagestyle{empty}
\begin{landscape}
\setcounter{table}{3}
\begin{center}
\begin{tabular}{llllllcrlll}
\multicolumn{10}{l}{{\bf {Table 3}} (continued).}\\
\hline
Designation & RA & Dec & Time & Identification
& Type & Counts & ${\chi}^{2}$/dof & $\lambda$ & F & D \\
            & (2000) & (2000) &      &              
&       & (s$^{-1}$\,half$^{-1}$) &  &  &  & (arc min) \\
\hline
EXMS B0534-695  & 05 34 34 & -69 30 43 & 1984/096 02:02 &       LMC X-1      &  XRB &    30.8\,$\pm$  4.7 &    1.01/ 1 &    22.0 & 4 & 31.7 \\
EXMS B0544-698C & 05 43 53 & -69 52 21 & 1984/104 16:49 &       LMC X-1      &  XRB &    28.7\,$\pm$  3.7 &    0.31/ 1 &    30.5 & 4 & 18.1 \\
EXMS B0545-700  & 05 45 05 & -70 03 35 & 1984/105 05:40 &       LMC X-1      &  XRB &    37.0\,$\pm$  6.5 &    0.40/ 1 &    17.5 & 4 & 29.0 \\
EXMS B0539-696  & 05 39 28 & -69 40 01 & 1984/106 05:26 &       LMC X-1      &  XRB &    37.0\,$\pm$  2.5 &    3.56/ 4 &   113.4 & 4 &  4.3 \\
EXMS B0540-697B & 05 39 60 & -69 42 58 & 1984/112 07:36 &       LMC X-1      &  XRB &    29.0\,$\pm$  1.7 &    0.01/ 1 &   170.3 & 4 &  0.8 \\
\hline
\end{tabular}
\end{center}
\end{landscape}
\newpage
\thispagestyle{empty}
\begin{landscape}
\setcounter{table}{3}
\begin{center}
\begin{tabular}{llllllcrlll}
\multicolumn{10}{l}{{\bf {Table 3}} (continued).}\\
\hline
Designation & RA & Dec & Time & Identification
& Type & Counts & ${\chi}^{2}$/dof & $\lambda$ & F & D \\
            & (2000) & (2000) &      &              
&       & (s$^{-1}$\,half$^{-1}$) &  &  &  & (arc min) \\
\hline
EXMS B0541-699B & 05 41 18 & -69 53 50 & 1985/312 22:20 &       LMC X-1      &  XRB &    27.8\,$\pm$  2.0 &   14.08/11 &   116.0 & 4 & 12.5 \\
EXMS B0543-699  & 05 42 44 & -69 53 24 & 1985/314 11:33 &      LMC X-1       &  XRB &    35.3\,$\pm$  3.3 &    2.26/ 5 &    58.8 & 3 & 39.7 \\
EXMS B0537-695  & 05 37 25 & -69 29 59 & 1985/320 16:39 &       LMC X-1      &  XRB &    33.4\,$\pm$  2.2 &    4.63/11 &   120.5 & 4 & 18.9 \\
EXMS B0543-700C & 05 43 17 & -70 04 45 & 1985/333 16:20 &      LMC X-1       &  XRB &    48.0\,$\pm$  5.2 &    3.88/11 &    43.4 & 3 & 26.1 \\
EXMS B0540-697F & 05 39 43 & -69 42 56 & 1985/352 16:39 &       LMC X-1      &  XRB &    33.6\,$\pm$  1.8 &    6.87/11 &   183.8 & 4 &  1.7 \\
\hline
\end{tabular}
\end{center}
\end{landscape}
\newpage
\thispagestyle{empty}
\begin{landscape}
\setcounter{table}{3}
\begin{center}
\begin{tabular}{llllllcrlll}
\multicolumn{10}{l}{{\bf {Table 3}} (continued).}\\
\hline
Designation & RA & Dec & Time & Identification
& Type & Counts & ${\chi}^{2}$/dof & $\lambda$ & F & D \\
            & (2000) & (2000) &      &              
&       & (s$^{-1}$\,half$^{-1}$) &  &  &  & (arc min) \\
\hline
EXMS B0751-678  & 07 51 50 & -67 58 47 & 1986/021 03:20 &      EXO0748-676   &  XRB &     9.6\,$\pm$  1.8 &   18.85/11 &    15.2 & 3 & 11.6 \\
EXMS B0749-679  & 07 49 14 & -68 04 47 & 1985/042 07:08 &       IPC075141-68 &  XRA &    75.3\,$\pm$  3.8 &   17.41/ 5 &   201.5 & 4 & 17.4 \\
EXMS B0818-425  & 08 20 23 & -42 42 52 & 1984/137 08:40 &   J082158.2-430022 &  RAS &    18.8\,$\pm$  3.3 &    0.00/ 1 &    16.6 & 4 & 23.8 \\
EXMS B0820-427  & 08 21 57 & -42 55 20 & 1984/323 19:33 &   J082158.2-430022 &  RAS &    18.3\,$\pm$  2.3 &    0.65/ 1 &    35.5 & 4 &  3.3 \\
EXMS B0818-435  & 08 20 24 & -43 41 02 & 1985/139 10:11 &   J082158.2-430022 &  RAS &    49.9\,$\pm$ 14.2 &    7.84/10 &     6.4 & 4 & 43.8 \\
\hline
\end{tabular}
\end{center}
\end{landscape}
\newpage
\thispagestyle{empty}
\begin{landscape}
\setcounter{table}{3}
\begin{center}
\begin{tabular}{llllllcrlll}
\multicolumn{10}{l}{{\bf {Table 3}} (continued).}\\
\hline
Designation & RA & Dec & Time & Identification
& Type & Counts & ${\chi}^{2}$/dof & $\lambda$ & F & D \\
            & (2000) & (2000) &      &              
&       & (s$^{-1}$\,half$^{-1}$) &  &  &  & (arc min) \\
\hline
EXMS B0918-548  & 09 19 58 & -55 04 30 & 1983/352 16:38 &       4U0918-549   &  XRB &    18.1\,$\pm$  3.3 &    0.76/ 1 &    15.3 & 4 &  8.7 \\
EXMS B0918-549C & 09 19 57 & -55 08 24 & 1984/083 14:51 &       4U0918-549   &  XRB &    34.7\,$\pm$  2.3 &    4.59/ 5 &   116.0 & 4 &  6.1 \\
EXMS B0918-549A & 09 19 48 & -55 09 42 & 1984/083 10:03 &       4U0918-549   &  XRB &    30.5\,$\pm$  2.4 &    0.55/ 4 &    84.2 & 4 &  6.2 \\
EXMS B0918-549B & 09 19 51 & -55 07 32 & 1984/083 14:49 &       4U0918-549   &  XRB &    30.6\,$\pm$  2.4 &    8.16/ 5 &    80.7 & 4 &  7.5 \\
EXMS B0919-551  & 09 21 04 & -55 18 58 & 1984/159 23:10 &       4U0918-549   &  XRB &    14.6\,$\pm$  2.0 &    3.02/ 4 &    26.5 & 4 &  2.5 \\
\hline
\end{tabular}
\end{center}
\end{landscape}
\newpage
\thispagestyle{empty}
\begin{landscape}
\setcounter{table}{3}
\begin{center}
\begin{tabular}{llllllcrlll}
\multicolumn{10}{l}{{\bf {Table 3}} (continued).}\\
\hline
Designation & RA & Dec & Time & Identification
& Type & Counts & ${\chi}^{2}$/dof & $\lambda$ & F & D \\
            & (2000) & (2000) &      &              
&       & (s$^{-1}$\,half$^{-1}$) &  &  &  & (arc min) \\
\hline
EXMS B1124-601B & 11 26 46 & -60 28 22 & 1984/054 05:38 &       Cen X-3      &  XRB &   116.1\,$\pm$ 10.5 &   13.88/11 &    60.9 & 4 & 41.7 \\
EXMS B1117-608  & 11 19 35 & -61 05 08 & 1984/161 08:32 &       Cen X-3      &  XRB &    44.8\,$\pm$  5.2 &    5.29/ 5 &    37.5 & 4 & 29.7 \\
EXMS B1119-603  & 11 21 26 & -60 35 42 & 1984/183 18:25 &       Cen X-3      &  XRB &    86.3\,$\pm$  1.8 &   95.89/12 &  1215.0 & 4 &  2.3 \\
EXMS B1120-601A & 11 22 18 & -60 27 39 & 1984/187 18:53 &      Cen X-3       &  XRB &   129.2\,$\pm$  2.0 &  123.39/11 &  2036.6 & 3 & 11.5 \\
EXMS B1119-601  & 11 21 56 & -60 26 22 & 1984/187 15:46 &      Cen X-3       &  XRB &   150.0\,$\pm$  2.1 &   79.27/11 &  2473.3 & 3 & 11.2 \\
\hline
\end{tabular}
\end{center}
\end{landscape}
\newpage
\thispagestyle{empty}
\begin{landscape}
\setcounter{table}{3}
\begin{center}
\begin{tabular}{llllllcrlll}
\multicolumn{10}{l}{{\bf {Table 3}} (continued).}\\
\hline
Designation & RA & Dec & Time & Identification
& Type & Counts & ${\chi}^{2}$/dof & $\lambda$ & F & D \\
            & (2000) & (2000) &      &              
&       & (s$^{-1}$\,half$^{-1}$) &  &  &  & (arc min) \\
\hline
EXMS B1227+125  & 12 30 18 & +12 15 50 & 1985/176 04:55 &      Virgo         &  CLU &    17.3\,$\pm$  1.5 &   15.10/11 &    65.3 & 3 &  4.1 \\
EXMS B1228+127B & 12 31 23 & +12 26 06 & 1985/355 14:30 &      Virgo         &  CLU &    24.5\,$\pm$  2.1 &    5.67/11 &    66.1 & 3 &  8.8 \\
EXMS B1228+128A & 12 31 02 & +12 31 56 & 1986/009 09:19 &      Virgo         &  CLU &    19.0\,$\pm$  1.6 &   10.90/11 &    66.8 & 3 &  8.0 \\
EXMS B1228+128B & 12 30 57 & +12 33 47 & 1986/009 22:49 &      Virgo         &  CLU &    19.4\,$\pm$  1.3 &   22.01/11 &   119.7 & 3 & 10.3 \\
EXMS B1235+708  & 12 37 34 & +70 36 56 & 1983/315 09:43 &   J123137.5+704417 &  RAS &    10.8\,$\pm$  3.1 &    3.07/ 5 &     6.0 & 2 & 24.0 \\
\hline
\end{tabular}
\end{center}
\end{landscape}
\newpage
\thispagestyle{empty}
\begin{landscape}
\setcounter{table}{3}
\begin{center}
\begin{tabular}{llllllcrlll}
\multicolumn{10}{l}{{\bf {Table 3}} (continued).}\\
\hline
Designation & RA & Dec & Time & Identification
& Type & Counts & ${\chi}^{2}$/dof & $\lambda$ & F & D \\
            & (2000) & (2000) &      &              
&       & (s$^{-1}$\,half$^{-1}$) &  &  &  & (arc min) \\
\hline
EXMS B1259+278B & 13 01 35 & +27 35 58 & 1985/175 12:53 &      Coma          &  CLU &    23.2\,$\pm$  5.3 &    1.28/ 5 &     9.8 & 3 & 31.2 \\
EXMS B1257+279  & 13 00 14 & +27 43 49 & 1985/176 04:31 &      Coma          &  CLU &    19.0\,$\pm$  1.8 &   13.25/11 &    54.0 & 3 & 11.5 \\
EXMS B1254+285  & 12 57 08 & +28 16 08 & 1985/355 14:08 &      Coma          &  CLU &    34.5\,$\pm$  6.4 &    6.08/11 &    15.7 & 3 & 38.8 \\
EXMS B1257+287A & 13 00 17 & +28 28 27 & 1986/014 09:26 &      Coma          &  CLU &    22.4\,$\pm$  3.1 &    5.59/11 &    25.6 & 3 & 30.2 \\
EXMS B1257+287B & 13 00 23 & +28 28 52 & 1986/014 23:07 &      Coma          &  CLU &    29.4\,$\pm$  2.6 &   17.08/11 &    62.8 & 3 & 29.9 \\
\hline
\end{tabular}
\end{center}
\end{landscape}
\newpage
\thispagestyle{empty}
\begin{landscape}
\setcounter{table}{3}
\begin{center}
\begin{tabular}{llllllcrlll}
\multicolumn{10}{l}{{\bf {Table 3}} (continued).}\\
\hline
Designation & RA & Dec & Time & Identification
& Type & Counts & ${\chi}^{2}$/dof & $\lambda$ & F & D \\
            & (2000) & (2000) &      &              
&       & (s$^{-1}$\,half$^{-1}$) &  &  &  & (arc min) \\
\hline
EXMS B1539-519  & 15 43 02 & -52 05 28 & 1985/226 04:25 &       4U1538-522   &  XRB &     7.7\,$\pm$  2.0 &    5.35/11 &     7.5 & 4 & 18.3 \\
EXMS B1545-627  & 15 50 17 & -62 56 33 & 1984/247 18:20 &   J154659.6-624741 &  RAS &    98.4\,$\pm$ 13.6 &    1.24/ 1 &    26.3 & 4 & 19.3 \\
EXMS B1540-623  & 15 44 41 & -62 33 23 & 1984/059 19:16 &         4U1543-624 &  XRB &    94.0\,$\pm$  7.4 &    0.37/ 2 &    82.1 & 3 & 22.4 \\
EXMS B1543-624A & 15 47 34 & -62 34 18 & 1984/061 00:46 &       4U1543-624   &  XRB &    68.6\,$\pm$  1.9 &    6.93/ 5 &   623.4 & 4 &  0.2 \\
EXMS B1543-624B & 15 47 36 & -62 38 44 & 1984/249 20:56 &         4U1543-624 &  XRB &    62.9\,$\pm$  3.9 &    1.05/ 1 &   139.5 & 3 &  1.8 \\
\hline
\end{tabular}
\end{center}
\end{landscape}
\newpage
\thispagestyle{empty}
\begin{landscape}
\setcounter{table}{3}
\begin{center}
\begin{tabular}{llllllcrlll}
\multicolumn{10}{l}{{\bf {Table 3}} (continued).}\\
\hline
Designation & RA & Dec & Time & Identification
& Type & Counts & ${\chi}^{2}$/dof & $\lambda$ & F & D \\
            & (2000) & (2000) &      &              
&       & (s$^{-1}$\,half$^{-1}$) &  &  &  & (arc min) \\
\hline
EXMS B1614-154  & 16 17 46 & -15 37 14 & 1984/055 10:14 &      Sco X-1       &  XRB &  4081.4\,$\pm$ 29.6 & 2333.61/ 1 & 10527.1 & 3 & 28.7 \\
EXMS B1620-156  & 16 23 15 & -15 45 39 & 1984/055 10:14 &      Sco X-1       &  XRB & 13432.2\,$\pm$571.0 &    0.16/ 1 &   289.1 & 3 & 28.7 \\
EXMS B1618-153  & 16 21 03 & -15 30 43 & 1984/243 10:09 &      Sco X-1       &  XRB &  2388.9\,$\pm$ 24.4 & 2398.68/ 1 &  5135.2 & 3 & 14.9 \\
EXMS B1618-156  & 16 20 57 & -15 46 49 & 1984/243 05:43 &      Sco X-1       &  XRB &  2001.8\,$\pm$ 22.7 & 2953.27/ 1 &  3885.5 & 3 & 16.0 \\
EXMS B1613-152  & 16 16 11 & -15 19 41 & 1986/053 15:07 &      Sco X-1       &  XRB &  3385.7\,$\pm$160.9 &    8.21/11 &   221.7 & 3 & 46.6 \\
\hline
\end{tabular}
\end{center}
\end{landscape}
\newpage
\thispagestyle{empty}
\begin{landscape}
\setcounter{table}{3}
\begin{center}
\begin{tabular}{llllllcrlll}
\multicolumn{10}{l}{{\bf {Table 3}} (continued).}\\
\hline
Designation & RA & Dec & Time & Identification
& Type & Counts & ${\chi}^{2}$/dof & $\lambda$ & F & D \\
            & (2000) & (2000) &      &              
&       & (s$^{-1}$\,half$^{-1}$) &  &  &  & (arc min) \\
\hline
EXMS B1642-452B & 16 46 08 & -45 17 31 & 1985/227 10:45 &      GX340+0       &  XRB &   570.3\,$\pm$  2.8 &   46.27/11 & 21393.2 & 3 & 13.9 \\
EXMS B1642-452A & 16 46 03 & -45 17 30 & 1985/227 10:45 &      GX340+0       &  XRB &   592.3\,$\pm$  4.2 &   38.57/11 &  9907.2 & 3 & 13.9 \\
EXMS B1642-454  & 16 45 46 & -45 31 56 & 1985/231 09:22 &       GX340+0      &  XRB &   397.2\,$\pm$  2.7 &   19.82/11 & 10476.6 & 4 &  4.8 \\
EXMS B1650-594  & 16 54 32 & -59 31 42 & 1985/069 06:18 &   J165259.4-594302 &  RAS &    21.5\,$\pm$  4.4 &   10.89/11 &    11.7 & 4 & 33.8 \\
EXMS B1646-596  & 16 51 10 & -59 44 38 & 1985/072 04:28 &   J165259.4-594302 &  RAS &     6.3\,$\pm$  1.7 &   15.21/11 &     7.1 & 4 &  3.8 \\
\hline
\end{tabular}
\end{center}
\end{landscape}
\newpage
\thispagestyle{empty}
\begin{landscape}
\setcounter{table}{3}
\begin{center}
\begin{tabular}{llllllcrlll}
\multicolumn{10}{l}{{\bf {Table 3}} (continued).}\\
\hline
Designation & RA & Dec & Time & Identification
& Type & Counts & ${\chi}^{2}$/dof & $\lambda$ & F & D \\
            & (2000) & (2000) &      &              
&       & (s$^{-1}$\,half$^{-1}$) &  &  &  & (arc min) \\
\hline
EXMS B1705-441  & 17 09 02 & -44 12 49 & 1984/254 19:01 &         4U1705-440 &  XRB &   130.2\,$\pm$  6.7 &    3.21/ 1 &   191.3 & 3 &  2.1 \\
EXMS B1705-401  & 17 08 50 & -40 14 57 & 1985/069 05:51 &   J170849.0-400910 &  RAS &    17.9\,$\pm$  4.0 &    8.38/11 &     9.9 & 4 & 33.8 \\
EXMS B1708+788  & 17 05 04 & +78 49 33 & 1984/313 16:14 &      A2256         &  CLU &     6.1\,$\pm$  1.3 &    5.50/ 4 &    10.5 & 3 &  1.5 \\
EXMS B1706+787  & 17 03 19 & +78 41 21 & 1984/314 15:32 &      A2256         &  CLU &     5.5\,$\pm$  0.9 &    7.99/12 &    19.6 & 3 &  0.2 \\
EXMS B1700+787  & 16 57 51 & +78 41 59 & 1985/093 20:49 &      A2256         &  CLU &     6.6\,$\pm$  1.6 &    2.30/ 5 &     8.9 & 3 & 15.7 \\
\hline
\end{tabular}
\end{center}
\end{landscape}
\newpage
\thispagestyle{empty}
\begin{landscape}
\setcounter{table}{3}
\begin{center}
\begin{tabular}{llllllcrlll}
\multicolumn{10}{l}{{\bf {Table 3}} (continued).}\\
\hline
Designation & RA & Dec & Time & Identification
& Type & Counts & ${\chi}^{2}$/dof & $\lambda$ & F & D \\
            & (2000) & (2000) &      &              
&       & (s$^{-1}$\,half$^{-1}$) &  &  &  & (arc min) \\
\hline
EXMS B1715-320  & 17 18 49 & -32 07 43 & 1985/069 05:37 &       M1715-321    &  XRB &    38.7\,$\pm$  4.0 &   10.45/11 &    47.6 & 4 & 33.8 \\
EXMS B1715-325  & 17 18 46 & -32 34 40 & 1985/231 10:58 &       M1715-321    &  XRB &    38.9\,$\pm$  7.9 &   10.82/11 &    14.5 & 4 & 24.0 \\
EXMS B1714-321B & 17 17 40 & -32 11 44 & 1985/256 10:52 &       M1715-321    &  XRB &    16.1\,$\pm$  1.1 &   18.07/10 &   104.2 & 4 & 14.0 \\
EXMS B1723-307  & 17 26 35 & -30 46 27 & 1983/257 16:28 &       E1724-307    &  XRB &    14.3\,$\pm$  2.7 &    1.68/ 4 &    14.5 & 4 & 12.6 \\
EXMS B1724-312  & 17 27 59 & -31 19 25 & 1984/073 06:12 &       E1724-307    &  XRB &    26.3\,$\pm$  4.2 &    3.07/11 &    20.0 & 4 & 31.7 \\
\hline
\end{tabular}
\end{center}
\end{landscape}
\newpage
\thispagestyle{empty}
\begin{landscape}
\setcounter{table}{3}
\begin{center}
\begin{tabular}{llllllcrlll}
\multicolumn{10}{l}{{\bf {Table 3}} (continued).}\\
\hline
Designation & RA & Dec & Time & Identification
& Type & Counts & ${\chi}^{2}$/dof & $\lambda$ & F & D \\
            & (2000) & (2000) &      &              
&       & (s$^{-1}$\,half$^{-1}$) &  &  &  & (arc min) \\
\hline
EXMS B1734-155  & 17 37 48 & -15 32 57 & 1983/269 23:48 &   J173735.1-152357 &  RAS &     5.9\,$\pm$  1.4 &    6.09/11 &     8.6 & 4 &  8.9 \\
EXMS B1733-270  & 17 36 17 & -27 03 32 & 1985/073 22:29 &       SL1735-269   &  XRB &    11.9\,$\pm$  2.0 &   21.97/11 &    17.6 & 4 & 26.2 \\
EXMS B1738-444  & 17 41 42 & -44 30 56 & 1986/074 11:05 &         4U1735-444 &  XRB &   132.6\,$\pm$  6.1 &    4.42/11 &   235.2 & 3 & 29.3 \\
EXMS B1735-445  & 17 39 06 & -44 32 08 & 1984/262 07:18 &   1WGA J1739.2-440 &  XRA &   529.4\,$\pm$ 10.9 &    3.34/ 5 &  1172.6 & 4 & 30.1 \\
EXMS B1738-288  & 17 41 33 & -28 55 21 & 1986/074 11:27 &   1WGA J1742.7-285 &  XRA &   375.1\,$\pm$  6.2 &   15.76/11 &  1832.1 & 4 & 15.0 \\
\hline
\end{tabular}
\end{center}
\end{landscape}
\newpage
\thispagestyle{empty}
\begin{landscape}
\setcounter{table}{3}
\begin{center}
\begin{tabular}{llllllcrlll}
\multicolumn{10}{l}{{\bf {Table 3}} (continued).}\\
\hline
Designation & RA & Dec & Time & Identification
& Type & Counts & ${\chi}^{2}$/dof & $\lambda$ & F & D \\
            & (2000) & (2000) &      &              
&       & (s$^{-1}$\,half$^{-1}$) &  &  &  & (arc min) \\
\hline
EXMS B1753-309  & 17 57 02 & -30 55 11 & 1985/231 11:08 &   J175721.2-304405 &  RAS &     4.6\,$\pm$  1.2 &   10.14/11 &     7.9 & 4 & 10.7 \\
EXMS B1754-308  & 17 57 45 & -30 52 48 & 1985/232 08:25 &   J175721.2-304405 &  RAS &    66.9\,$\pm$ 13.7 &   12.99/11 &    11.9 & 4 &  8.1 \\
EXMS B1752-338  & 17 55 38 & -33 52 33 & 1984/077 20:10 &       4U1755-338   &  XRB &   145.1\,$\pm$ 11.8 &    3.23/ 2 &    76.1 & 4 & 35.5 \\
EXMS B1754-338A & 17 57 60 & -33 52 03 & 1984/081 13:18 &         4U1755-338 &  XRB &   138.0\,$\pm$  6.0 &    7.01/ 2 &   269.2 & 3 &  8.4 \\
EXMS B1755-339A & 17 58 55 & -33 54 18 & 1984/245 19:52 &         4U1755-338 &  XRB &   172.2\,$\pm$  2.8 &   18.13/11 &  1957.8 & 3 &  6.0 \\
\hline
\end{tabular}
\end{center}
\end{landscape}
\newpage
\thispagestyle{empty}
\begin{landscape}
\setcounter{table}{3}
\begin{center}
\begin{tabular}{llllllcrlll}
\multicolumn{10}{l}{{\bf {Table 3}} (continued).}\\
\hline
Designation & RA & Dec & Time & Identification
& Type & Counts & ${\chi}^{2}$/dof & $\lambda$ & F & D \\
            & (2000) & (2000) &      &              
&       & (s$^{-1}$\,half$^{-1}$) &  &  &  & (arc min) \\
\hline
EXMS B1810+003  & 18 12 50 & +00 21 17 & 1984/270 13:12 &   1WGA J1814.5+001 &  XRA &    46.4\,$\pm$  0.4 &    1.63/ 5 &    36.3 & 4 & 26.0 \\
EXMS B1812-116  & 18 15 33 & -11 38 28 & 1983/260 11:48 &       M1812-12     &  XRB &    24.8\,$\pm$  6.3 &    0.02/ 1 &     8.5 & 4 & 26.7 \\
EXMS B1811-126  & 18 14 18 & -12 39 45 & 1984/249 21:45 &       M1812-12     &  XRB &    12.2\,$\pm$  5.3 &    5.75/11 &     2.7 & 4 & 35.2 \\
EXMS B1813-126  & 18 16 43 & -12 40 53 & 1984/249 21:45 &           M1812-12 &  XRB &    22.0\,$\pm$  5.3 &    4.02/11 &     9.5 & 3 & 35.2 \\
EXMS B1812-126  & 18 15 00 & -12 35 40 & 1984/251 04:06 &       M1812-12     &  XRB &    40.7\,$\pm$ 10.9 &    7.94/11 &     7.1 & 4 & 30.7 \\
\hline
\end{tabular}
\end{center}
\end{landscape}
\newpage
\thispagestyle{empty}
\begin{landscape}
\setcounter{table}{3}
\begin{center}
\begin{tabular}{llllllcrlll}
\multicolumn{10}{l}{{\bf {Table 3}} (continued).}\\
\hline
Designation & RA & Dec & Time & Identification
& Type & Counts & ${\chi}^{2}$/dof & $\lambda$ & F & D \\
            & (2000) & (2000) &      &              
&       & (s$^{-1}$\,half$^{-1}$) &  &  &  & (arc min) \\
\hline
EXMS B1837+050  & 18 39 43 & +05 03 03 & 1983/279 11:17 &       Ser X-1      &  XRB &   304.0\,$\pm$  6.7 &    0.62/ 1 &  1043.9 & 4 & 24.6 \\
EXMS B1837+051A & 18 40 09 & +05 10 41 & 1984/099 19:28 &      Ser X-1       &  XRB &   328.7\,$\pm$  5.3 &    4.57/ 5 &  1923.8 & 3 &  4.6 \\
EXMS B1837+049A & 18 40 00 & +04 58 59 & 1984/099 14:47 &       Ser X-1      &  XRB &   278.5\,$\pm$  4.5 &    2.77/ 4 &  1894.6 & 4 &  3.3 \\
EXMS B1837+049B & 18 39 60 & +04 57 45 & 1984/099 14:49 &       Ser X-1      &  XRB &   328.5\,$\pm$  5.0 &   13.36/ 5 &  2196.0 & 4 &  4.5 \\
EXMS B1837+051B & 18 40 02 & +05 09 48 & 1984/099 19:30 &      Ser X-1       &  XRB &   317.9\,$\pm$  5.1 &    1.96/ 4 &  1964.0 & 3 &  3.4 \\
\hline
\end{tabular}
\end{center}
\end{landscape}
\newpage
\thispagestyle{empty}
\begin{landscape}
\setcounter{table}{3}
\begin{center}
\begin{tabular}{llllllcrlll}
\multicolumn{10}{l}{{\bf {Table 3}} (continued).}\\
\hline
Designation & RA & Dec & Time & Identification
& Type & Counts & ${\chi}^{2}$/dof & $\lambda$ & F & D \\
            & (2000) & (2000) &      &              
&       & (s$^{-1}$\,half$^{-1}$) &  &  &  & (arc min) \\
\hline
EXMS B1906+095  & 19 08 34 & +09 37 09 & 1985/286 13:29 &       4U1907+097   &  XRB &     7.1\,$\pm$  1.7 &    7.34/11 &     9.1 & 4 & 17.5 \\
EXMS B1904+522  & 19 05 22 & +52 21 44 & 1985/171 20:15 &   J190826.1+522532 &  RAS &    10.1\,$\pm$  2.5 &   11.28/11 &     8.0 & 4 & 27.6 \\
EXMS B1905+008  & 19 08 04 & +00 54 38 & 1985/099 08:12 &   1WGA J1910.6+010 &  XRA &    62.8\,$\pm$  7.1 &    1.42/ 5 &    39.3 & 4 & 39.0 \\
EXMS B1907+075  & 19 09 45 & +07 34 57 & 1983/285 01:52 &       1H1907+074   &  XRA &     7.8\,$\pm$  1.9 &    9.88/11 &    10.7 & 4 & 15.2 \\
EXMS B1907+076A & 19 09 42 & +07 43 37 & 1984/099 11:32 &       1H1907+074   &  XRA &    10.6\,$\pm$  2.7 &    0.29/ 1 &     7.7 & 4 & 12.5 \\
\hline
\end{tabular}
\end{center}
\end{landscape}
\newpage
\thispagestyle{empty}
\begin{landscape}
\setcounter{table}{3}
\begin{center}
\begin{tabular}{llllllcrlll}
\multicolumn{10}{l}{{\bf {Table 3}} (continued).}\\
\hline
Designation & RA & Dec & Time & Identification
& Type & Counts & ${\chi}^{2}$/dof & $\lambda$ & F & D \\
            & (2000) & (2000) &      &              
&       & (s$^{-1}$\,half$^{-1}$) &  &  &  & (arc min) \\
\hline
EXMS B1919+440B & 19 21 03 & +44 10 11 & 1985/153 12:43 &      A2319         &  CLU &     9.2\,$\pm$  2.1 &    1.20/ 4 &    10.0 & 3 &  4.7 \\
EXMS B1917+441  & 19 18 38 & +44 13 25 & 1985/291 23:59 &      A2319         &  CLU &     7.6\,$\pm$  2.3 &    3.89/ 6 &     4.5 & 3 & 19.4 \\
EXMS B1919+442A & 19 20 50 & +44 18 57 & 1985/292 00:03 &      A2319         &  CLU &    10.2\,$\pm$  2.1 &    6.22/11 &    11.7 & 3 & 20.8 \\
EXMS B1919+442B & 19 20 58 & +44 22 56 & 1985/292 06:51 &      A2319         &  CLU &    10.2\,$\pm$  2.8 &   10.86/11 &     6.8 & 3 & 24.4 \\
EXMS B1920+442  & 19 22 14 & +44 20 14 & 1985/292 00:00 &      A2319         &  CLU &     8.5\,$\pm$  2.3 &    6.94/11 &     6.9 & 3 & 28.6 \\
\hline
\end{tabular}
\end{center}
\end{landscape}
\newpage
\thispagestyle{empty}
\begin{landscape}
\setcounter{table}{3}
\begin{center}
\begin{tabular}{llllllcrlll}
\multicolumn{10}{l}{{\bf {Table 3}} (continued).}\\
\hline
Designation & RA & Dec & Time & Identification
& Type & Counts & ${\chi}^{2}$/dof & $\lambda$ & F & D \\
            & (2000) & (2000) &      &              
&       & (s$^{-1}$\,half$^{-1}$) &  &  &  & (arc min) \\
\hline
EXMS B1956+113  & 19 59 02 & +11 31 05 & 1985/302 19:07 &         4U1957+115 &  XRB &    53.4\,$\pm$  1.7 &    9.16/11 &   515.9 & 3 &  8.1 \\
EXMS B1958+405  & 19 59 59 & +40 43 09 & 1983/295 04:20 &      Cygnus A      &  AGN &     3.8\,$\pm$  0.9 &    4.52/ 5 &     8.6 & 3 &  0.7 \\
EXMS B1957+404  & 19 58 45 & +40 35 23 & 1983/311 08:27 &      Cygnus A      &  AGN &     9.9\,$\pm$  1.7 &    7.14/11 &    18.0 & 3 & 10.7 \\
EXMS B1958+407  & 20 00 22 & +40 53 44 & 1983/314 16:16 &      Cygnus A      &  AGN &     5.4\,$\pm$  1.1 &    5.71/11 &    12.2 & 3 & 10.8 \\
EXMS B1956+404  & 19 57 58 & +40 35 49 & 1984/132 22:40 &      Cygnus A      &  AGN &     9.2\,$\pm$  1.3 &    5.50/11 &    25.8 & 3 & 16.7 \\
\hline
\end{tabular}
\end{center}
\end{landscape}
\newpage
\thispagestyle{empty}
\begin{landscape}
\setcounter{table}{3}
\begin{center}
\begin{tabular}{llllllcrlll}
\multicolumn{10}{l}{{\bf {Table 3}} (continued).}\\
\hline
Designation & RA & Dec & Time & Identification
& Type & Counts & ${\chi}^{2}$/dof & $\lambda$ & F & D \\
            & (2000) & (2000) &      &              
&       & (s$^{-1}$\,half$^{-1}$) &  &  &  & (arc min) \\
\hline
EXMS B2029+406  & 20 31 13 & +40 48 55 & 1984/326 00:25 &       Cyg X-3      &  XRB &   160.6\,$\pm$  3.2 &    5.84/ 5 &  1233.5 & 4 & 15.3 \\
EXMS B2030+407  & 20 31 50 & +40 56 51 & 1984/327 01:46 &      Cyg X-3       &  XRB &   209.1\,$\pm$  3.0 &    5.71/ 5 &  2411.1 & 3 &  5.4 \\
EXMS B2031+410  & 20 32 59 & +41 14 25 & 1985/137 18:48 &      Cyg X-3       &  XRB &   131.5\,$\pm$  2.9 &    7.39/11 &  1035.1 & 3 & 13.0 \\
EXMS B2030+410  & 20 32 10 & +41 13 05 & 1985/138 20:38 &      Cyg X-3       &  XRB &   133.0\,$\pm$  2.6 &    1.65/ 4 &  1291.8 & 3 &  2.8 \\
EXMS B2029+409A & 20 31 44 & +41 08 46 & 1985/138 09:32 &      Cyg X-3       &  XRB &   115.9\,$\pm$  2.4 &    0.81/ 5 &  1143.7 & 3 &  0.5 \\
\hline
\end{tabular}
\end{center}
\end{landscape}
\newpage
\thispagestyle{empty}
\begin{landscape}
\setcounter{table}{3}
\begin{center}
\begin{tabular}{llllllcrlll}
\multicolumn{10}{l}{{\bf {Table 3}} (continued).}\\
\hline
Designation & RA & Dec & Time & Identification
& Type & Counts & ${\chi}^{2}$/dof & $\lambda$ & F & D \\
            & (2000) & (2000) &      &              
&       & (s$^{-1}$\,half$^{-1}$) &  &  &  & (arc min) \\
\hline
EXMS B2156-302  & 21 58 57 & -29 58 37 & 1985/145 19:22 &       PKS 2155-304 &  AGN &    11.2\,$\pm$  2.5 &   13.66/11 &    10.1 & 3 & 14.0 \\
EXMS B2204+546  & 22 06 25 & +54 51 03 & 1983/356 07:53 &       4U2206+543   &  XRB &    13.2\,$\pm$  2.4 &    4.39/11 &    14.6 & 4 & 23.9 \\
EXMS B2205+542  & 22 07 28 & +54 28 40 & 1984/182 00:40 &       4U2206+543   &  XRB &     7.6\,$\pm$  1.3 &    3.88/ 5 &    16.4 & 4 &  4.4 \\
EXMS B2206+542  & 22 07 55 & +54 28 42 & 1984/185 12:06 &       4U2206+543   &  XRB &     7.7\,$\pm$  0.9 &    8.25/11 &    32.7 & 4 &  2.0 \\
EXMS B2207+543B & 22 09 43 & +54 34 10 & 1984/243 06:41 &       4U2206+543   &  XRB &    22.9\,$\pm$  2.8 &    7.52/ 5 &    34.5 & 4 & 15.7 \\
\hline
\end{tabular}
\end{center}

\end{landscape}
\newpage

\begin{landscape}
\thispagestyle{empty}
\begin{center}
\begin{tabular}{llllllcrl}
\multicolumn{9}{l}{{\bf {Table 4.}} Multiply-identified EXMS entries}\\
\hline
Designation & RA & Dec & Time & Identification
& Type & Counts & ${\chi}^{2}$/dof & $\lambda$ \\
            & (2000) & (2000)  &                     
&            &      & (s$^{-1}$\,half$^{-1}$) &  &   \\
\hline
EXMS B0016-722  & 00 18 45 & -71 56 46 & 1984/295 11:39 & J002404.6-720456               &  RAS &     9.5\,$\pm$  1.1 &   13.36/10 &    37.6 \\
& & & &J001726.8-715030              & & & &\\
EXMS B0053-752  & 00 55 03 & -74 56 41 & 1984/300 22:55 &          U Tuc, AY Tuc, CF Tuc &  VST &     4.8\,$\pm$  1.3 &   11.6/ 5.0 &     6.5 \\
EXMS B0114+005  & 01 16 50 & +00 47 31 & 1984/191 23:25 & J011704.2+000025               &  RAS &     5.4\,$\pm$  1.2 &    5.38/ 4 &     9.8 \\
& & & &J011504.5+002223              & & & &\\
\hline
\end{tabular}
\end{center}
\end{landscape}
\newpage
\begin{landscape}
\thispagestyle{empty}
\setcounter{table}{3}
\begin{center}
\begin{tabular}{llllllcrl}
\multicolumn{9}{l}{{\bf {Table 4}}(continued)}\\
\hline
Designation & RA & Dec & Time & Identification
& Type & Counts & ${\chi}^{2}$/dof & $\lambda$ \\
            & (2000) & (2000)       &               
&            &      & (s$^{-1}$\,half$^{-1}$) &  &   \\
\hline
EXMS B0557+228  & 06 00 15 & +22 53 50 & 1985/267 02:07 & J055729.1+230848               &  RAS &     4.8\,$\pm$  1.3 &    3.81/11 &     6.9 \\
& & & &J060205.7+231004              & & & &\\
EXMS B0658+753  & 07 04 41 & +75 13 39 & 1984/089 08:29 & J070406.9+751337               &  RAS &     6.2\,$\pm$  1.7 &    2.14/ 5 &     6.6 \\
& & & &J070415.5+752442              & & & &\\
EXMS B0834+254  & 08 37 57 & +25 16 06 & 1983/313 19:57 & J083811.0+245336               &  RAS &     4.6\,$\pm$  1.2 &    3.38/11 &     6.9 \\
\hline
\end{tabular}
\end{center}
\end{landscape}
\newpage
\begin{landscape}
\thispagestyle{empty}
\setcounter{table}{3}
\begin{center}
\begin{tabular}{llllllcrl}
\multicolumn{9}{l}{{\bf {Table 4}}(continued)}\\
\hline
Designation & RA & Dec & Time & Identification
& Type & Counts & ${\chi}^{2}$/dof & $\lambda$ \\
            & (2000) & (2000)       &               
&            &      & (s$^{-1}$\,half$^{-1}$) &  &   \\
\hline
EXMS B1743-303A & 17 47 02 & -30 24 07 & 1984/106 19:31 &        SL1744-300,  SL1744-299 &  XRB &    21.6\,$\pm$  1.8 &   16.3/10.0 &    73.1 \\
EXMS B1745-295  & 17 49 03 & -29 36 11 & 1984/077 20:12 &                 GC X-1, GC X-2 &  XRB &    16.1\,$\pm$  3.8 &    0.3/ 1.0 &     9.0 \\
EXMS B1806+681  & 18 06 46 & +68 12 07 & 1984/253 21:00 & IPC180422+67, MS1804.3+675     &  XRA &     8.7\,$\pm$  2.5 &    0.22/ 1 &     6.2 \\
& & & &IPC180026+68                  & & & &\\
EXMS B1820-308A & 18 23 34 & -30 49 55 & 1984/106 19:42 & HRI182027-30, HRI182027-30     &  XRA &   179.6\,$\pm$  2.8 &   22.14/11 &  2024.2 \\
\hline
\end{tabular}
\end{center}

\end{landscape}

\begin{landscape}
\thispagestyle{empty}
\begin{center}
\begin{tabular}{llllcccccrl}
\multicolumn{11}{l}{{\bf {Table 5.}} Uncertainty regions for unidentified EXMS entries}\\
\hline
Designation & RA & Dec & Time
& C1 & C2 & C3 & C4 & Counts & ${\chi}^{2}$/dof & $\lambda$ \\
     & (2000) & (2000) &    
&    &    &    &    & (s$^{-1}$\,half$^{-1}$) &   &    \\
\hline
EXMS B0029-841  & 23 57 48 & -84 16 31 & 1984/266 08:46 &   0.909,  -83.78 &  14.862,  -84.05 &  14.661,  -84.51 & 359.452,  -84.28 &    10.5\,$\pm$  2.4 &    1.69/ 3 &     9.4 \\
EXMS B0037+673  & 00 42 25 & +67 54 04 & 1984/040 22:10 &  10.927,   67.77 &   8.181,   66.71 &   7.857,   66.84 &  10.603,   67.90 &     5.8\,$\pm$  1.2 &   11.50/11 &    11.1 \\
EXMS B0134+380  & 01 32 30 & +38 40 09 & 1985/218 05:35 &  23.303,   38.74 &  24.171,   37.41 &  23.995,   37.33 &  23.125,   38.67 &     9.5\,$\pm$  1.8 &    3.18/ 5 &    14.8 \\
EXMS B0143+582  & 01 44 51 & +58 58 36 & 1983/358 12:18 &  26.109,   58.98 &  25.748,   57.50 &  25.850,   57.49 &  26.214,   58.98 &    48.3\,$\pm$  3.2 &    0.82/ 2 &   155.1 \\
EXMS B0216+571  & 02 15 17 & +56 24 28 & 1983/359 18:13 &  33.732,   56.42 &  34.335,   57.88 &  34.428,   57.87 &  33.821,   56.41 &    42.6\,$\pm$  3.0 &    5.06/ 2 &   188.4 \\
\hline
\end{tabular}
\end{center}
\end{landscape}
\newpage
\begin{landscape}
\thispagestyle{empty}
\setcounter{table}{3}
\begin{center}
\begin{tabular}{llllcccccrl}
\multicolumn{11}{l}{{\bf {Table 5}}(continued)}\\
\hline
Designation & RA & Dec & Time
& C1 & C2 & C3 & C4 & Counts & ${\chi}^{2}$/dof & $\lambda$ \\
     & (2000) & (2000) &    
&    &    &    &    & (s$^{-1}$\,half$^{-1}$) &  &    \\
\hline
EXMS B0718+235  & 07 18 47 & +24 19 56 & 1983/282 14:54 & 109.918,   24.30 & 109.687,   22.82 & 109.475,   22.85 & 109.696,   24.33 &     5.0\,$\pm$  1.3 &   18.36/11 &     7.4 \\
EXMS B0742-688  & 07 43 03 & -68 03 49 & 1984/198 13:19 & 115.572,  -68.06 & 115.542,  -69.56 & 115.747,  -69.56 & 115.763,  -68.06 &    23.1\,$\pm$  2.4 &    7.11/ 2 &    46.2 \\
EXMS B0744-287  & 07 47 19 & -28 56 09 & 1984/303 12:29 & 116.838,  -28.90 & 115.187,  -28.51 & 115.176,  -28.54 & 116.828,  -28.94 &    46.3\,$\pm$  3.7 &    0.74/ 1 &    86.4 \\
EXMS B0801-670  & 07 59 14 & -66 21 49 & 1984/365 06:48 & 119.897,  -66.35 & 121.193,  -67.76 & 121.101,  -67.78 & 119.810,  -66.36 &    31.1\,$\pm$  1.6 &    4.83/ 5 &   194.6 \\
EXMS B0803-687  & 08 02 20 & -68 00 36 & 1984/198 16:52 & 120.687,  -68.01 & 121.015,  -69.50 & 120.902,  -69.51 & 120.581,  -68.01 &    21.2\,$\pm$  1.1 &   13.57/11 &   191.4 \\
\hline
\end{tabular}
\end{center}
\end{landscape}
\newpage
\begin{landscape}
\thispagestyle{empty}
\setcounter{table}{3}
\begin{center}
\begin{tabular}{llllcccccrl}
\multicolumn{11}{l}{{\bf {Table 5}}(continued)}\\
\hline
Designation & RA & Dec & Time
& C1 & C2 & C3 & C4 & Counts & ${\chi}^{2}$/dof & $\lambda$ \\
     & (2000) & (2000) &    
&    &    &    &    & (s$^{-1}$\,half$^{-1}$) &  &    \\
\hline
EXMS B1550-461  & 15 54 21 & -46 19 43 & 1986/056 10:40 & 238.620,  -46.23 & 236.524,  -45.87 & 236.487,  -45.96 & 238.587,  -46.33 &     9.4\,$\pm$  1.2 &   12.45/11 &    30.5 \\
EXMS B1550-609  & 15 56 33 & -61 14 49 & 1984/247 18:19 & 239.224,  -61.09 & 236.274,  -60.66 & 236.173,  -60.81 & 239.136,  -61.25 &    10.3\,$\pm$  2.2 &    0.00/ 1 &    11.2 \\
EXMS B1610-605  & 16 04 46 & -60 20 01 & 1984/249 20:57 & 241.123,  -60.47 & 244.115,  -60.79 & 244.169,  -60.65 & 241.190,  -60.33 &    13.0\,$\pm$  2.9 &    0.26/ 2 &    10.1 \\
EXMS B1618-358  & 16 14 30 & -35 51 12 & 1984/244 02:37 & 243.675,  -35.63 & 245.497,  -35.88 & 245.454,  -36.10 & 243.626,  -35.85 &     6.5\,$\pm$  1.8 &    2.44/ 4 &     6.5 \\
EXMS B1618-651  & 16 21 24 & -64 24 31 & 1985/070 00:55 & 245.068,  -64.37 & 244.080,  -65.82 & 244.374,  -65.85 & 245.348,  -64.41 &     4.8\,$\pm$  0.7 &   10.50/ 9 &    22.4 \\
\hline
\end{tabular}
\end{center}
\end{landscape}
\newpage
\begin{landscape}
\thispagestyle{empty}
\setcounter{table}{3}
\begin{center}
\begin{tabular}{llllcccccrl}
\multicolumn{11}{l}{{\bf {Table 5}}(continued)}\\
\hline
Designation & RA & Dec & Time
& C1 & C2 & C3 & C4 & Counts & ${\chi}^{2}$/dof & $\lambda$ \\
     & (2000) & (2000) &    
&    &    &    &    & (s$^{-1}$\,half$^{-1}$) &  &    \\
\hline
EXMS B1755+657  & 17 48 21 & +65 54 15 & 1984/247 04:55 & 267.244,   66.12 & 270.729,   65.65 & 270.545,   65.44 & 267.086,   65.90 &     4.5\,$\pm$  1.2 &    6.18/ 5 &     6.5 \\
EXMS B1807-262  & 18 07 48 & -25 28 13 & 1983/268 02:46 & 271.906,  -25.47 & 271.927,  -26.97 & 271.970,  -26.97 & 271.950,  -25.47 &    50.1\,$\pm$  2.4 &    5.90/ 2 &   283.9 \\
EXMS B1809-365  & 18 09 40 & -35 46 34 & 1985/110 16:03 & 272.489,  -35.78 & 272.258,  -37.27 & 272.184,  -37.26 & 272.416,  -35.78 &    14.8\,$\pm$  1.1 &   13.58/11 &   106.6 \\
EXMS B1810-337  & 18 10 19 & -34 29 30 & 1984/246 03:26 & 272.847,  -34.47 & 272.686,  -32.98 & 272.421,  -33.00 & 272.578,  -34.49 &     4.8\,$\pm$  1.3 &   11.22/11 &     6.6 \\
EXMS B1827-100  & 18 30 35 & -09 55 27 & 1985/088 08:01 & 277.653,  -10.03 & 276.132,  -10.12 & 276.126,  -10.01 & 277.647,   -9.92 &     7.4\,$\pm$  1.0 &   12.27/11 &    30.0 \\
\hline
\end{tabular}
\end{center}
\end{landscape}
\newpage
\begin{landscape}
\thispagestyle{empty}
\setcounter{table}{3}
\begin{center}
\begin{tabular}{llllcccccrl}
\multicolumn{11}{l}{{\bf {Table 5}}(continued)}\\
\hline
Designation & RA & Dec & Time
& C1 & C2 & C3 & C4 & Counts & ${\chi}^{2}$/dof & $\lambda$ \\
     & (2000) & (2000) &    
&    &    &    &    & (s$^{-1}$\,half$^{-1}$) &  &    \\
\hline
EXMS B2121+328  & 21 19 32 & +33 25 07 & 1984/156 17:37 & 320.112,   33.53 & 321.027,   32.25 & 320.802,   32.13 & 319.885,   33.42 &     6.4\,$\pm$  1.7 &    2.18/ 5 &     7.0 \\
EXMS B2124+751  & 21 14 17 & +74 49 12 & 1984/243 06:24 & 318.281,   75.03 & 323.909,   75.48 & 324.129,   75.26 & 318.572,   74.82 &     6.2\,$\pm$  1.7 &    0.78/ 4 &     6.9 \\
EXMS B2127+662  & 21 31 26 & +66 54 23 & 1986/015 01:24 & 322.651,   66.94 & 321.143,   65.57 & 321.343,   65.54 & 322.860,   66.91 &     9.9\,$\pm$  1.2 &   35.10/11 &    34.4 \\
EXMS B2128+488  & 21 30 58 & +48 11 51 & 1984/168 00:12 & 322.967,   48.32 & 321.531,   49.48 & 321.306,   49.36 & 322.743,   48.20 &     5.2\,$\pm$  1.2 &    8.93/11 &     9.4 \\
EXMS B2135-356  & 21 35 21 & -34 51 28 & 1985/153 04:57 & 323.563,  -34.90 & 323.908,  -36.37 & 324.186,  -36.33 & 323.836,  -34.86 &     4.3\,$\pm$  1.2 &    6.32/11 &     6.5 \\
\hline
\end{tabular}
\end{center}

\end{landscape}
\end{document}